\documentclass[usenatbib]{mn2e}   
\pdfoutput=1 
\usepackage[pdftex]{graphicx} 
\usepackage{times}
\usepackage{rotate} 
\usepackage{amssymb} 
\usepackage{rotating}
\usepackage{epstopdf} 

\title[The physics of  AGN evolution]{Observational constraints on the
  physics   behind  the   evolution   of  AGN   since  $\bf   z\sim1$}
\author[Georgakakis et al.]   {A. Georgakakis$^{1}$, A.  L.  Coil$^2$,
  C.  N.   A. Willmer$^{3}$, K.  Nandra$^{4}$,  D. D.  Kocevski$^{5}$,
  \newauthor   M.   C.    Cooper$^6$\thanks{Hubble  Fellow},   D.   J.
  Rosario$^{4}$,   D.     C.    Koo$^{5}$,   J.     R.    Trump$^{5}$,
  S. Juneau$^{3}$ \\ \\ $^1$National Observatory of Athens, V.  Paulou
  \& I.  Metaxa, 11532,  Greece\\ $^2$Department of Physics and Center
  for Astrophysics  and Space Sciences, University  of California, San
  Diego,   9500  Gilman   Dr.,  La   Jolla,  CA   92093\\  $^3$Steward
  Observatory, University of Arizona, 933 North Cherry Avenue, Tucson,
  AZ 85721,  USA\\ $^4$Max Planck  Institut f\"{u}r Extraterrestrische
  Physik, Giessenbachstra\ss e, 85748 Garching, Germany\\ $^5$UCO/Lick
  Observatory, University of California, Santa Cruz, 1156 High Street,
  Santa Cruz,  CA 95064\\ $^6$Center for  Galaxy Evolution, Department
  of  Physics and  Astronomy, University  of California,  Irvine, 4129
  Frederick Reines Hall, Irvine, CA 92697, USA\\ }

\begin{document}
\maketitle

\begin{abstract}
We explore the  evolution with redshift of the  rest-frame colours and
space densities  of AGN  hosts (relative to  normal galaxies)  to shed
light on the dominant  mechanism that triggers accretion
onto supermassive black holes as a function of cosmic time.  Data from
serendipitous wide-area  XMM surveys of the  SDSS footprint (XMM/SDSS,
Needles  in  the  Haystack  survey)  are combined  with  Chandra  deep
observations  in the  AEGIS,  GOODS-North and  GOODS-South to  compile
uniformly selected samples of  moderate luminosity X-ray AGN ($L_X(\rm
2 - 10 \, keV ) = 10^{41} - 10^{44} \, erg\,s^{-1}$) at redshifts
0.1, 0.3 and 0.8.  It is found  that the fraction of AGN hosted by red
versus blue galaxies  does not change with redshift.   Also, the X-ray
luminosity  density  associated with  either  red  or  blue AGN  hosts
remains nearly constant since  $z=0.8$.  X-ray AGN represent a roughly
fixed  fraction of  the space  density  of galaxies  of given  optical
luminosity at  all redshifts probed  by our samples.  In  contrast the
fraction of X-ray AGN among galaxies of a given stellar mass decreases
with  decreasing  redshift.   These  findings suggest  that  the  same
process  or combination  of processes  for fueling  supermassive black
holes are in  operation in the last 5\,Gyrs of  cosmic time.  The data
are consistent with a picture in which the drop of the accretion power
during that period (1\,dex since $z=0.8$) is related to the decline of
the space density of available AGN hosts, as a result of the evolution
of the specific star-formation  rate of the overall galaxy population.
Scenarios  which attribute  the evolution  of moderate  luminosity AGN
since $z\approx1$ to changes in the suppermassive black hole accretion
mode are not favored by our results.
\end{abstract}
\begin{keywords} 
  galaxies: active -- galaxies: Seyferts -- X-rays: diffuse background
\end{keywords} 

\section{Introduction}\label{sec_intro}

Understanding  how galaxies assemble  their stellar  mass and  how the
supermassive  black holes (SMBH)  at their  centres grow  remain major
challenges  in current  astrophysical  research. Recent  observational
evidence  indicates  that the  two  processes  are intimately  related
\citep[e.g.][]{Ferrarese2000,  Gebhardt2000} and  therefore  should be
studied in conjunction: SMBHs affect  the evolution of their hosts and
vice versa.  Indeed, many modern semi-analytic models (SAMs) of galaxy
formation   include   prescriptions    for   the   growth   of   SMBHs
\citep[e.g.][]{Wang2007,  Monaco2007,  Somerville2008}.   In  most  of
those simulations it is assumed  that SMBHs, when active, regulate the
formation of new  stars through some form of  as yet poorly understood
feedback  process, thereby  affecting the  evolutionary path  of their
hosts.  Without  the energy input  from Active Galactic  Nuclei (AGN),
many SAMs  fail to reproduce fundamental properties  of galaxies, such
as their  colour bimodality \citep[e.g.][]{Cattaneo2007, Cattaneo2009}
and  luminosity  function  \citep[e.g.][]{Benson2003,Fontanot2007_lf}.
As a result of the development of AGN/galaxy co-evolution models it is
now possible to  have a physical description of  the accretion history
of  the Universe,  which  observationally, shows  a  broad plateau  at
$z\approx1-4$  followed by  a rapid  decline by  almost 1\,dex  to the
present day \citep[e.g.][]{Aird2010}.

The mechanisms adopted  by SAMs to trigger AGN  activity include major
gaseous  mergers   \citep[e.g.][]{DiMatteo2005,  Cen2011},  stochastic
accretion  of cold gas  via disk  instabilities or  minor interactions
\citep{Hopkins_Hernquist2006}  and quiescent inflows  of hot  gas from
the  galaxy halo  \citep{Croton2006}.   Hydrodynamic simulations  also
propose that the secular evolution of galaxies provides a large enough
reservoir of gas through stellar  winds, which could give episodic AGN
activity       and      substantial       black       hole      growth
\citep[e.g.][]{Ciotti_Ostriker1997,   Ciotti_Ostriker2007}.    To  our
knowledge this  process has not as  yet been implemented  in SAMs. The
different SMBH  fueling modes above operate by  design under different
conditions and therefore make  specific predictions for the properties
of the  galaxies that host  AGN, including their  star-formation rate,
morphology and environment.  This offers an observational tool to test
different SMBH growth models.

The evidence above has motivated efforts to study AGN hosts using data
from   large    multiwavelength   programs,   such    as   the   AEGIS
\citep[All-wavelength     Extended    Groth     Strip    International
  Survey;][]{Davis2007},    the    COSMOS   \citep[Cosmic    Evolution
  Survey;][]{Scoville2007},  the  GOODS  (Great Observatories  Origins
Survey)  and  the  XBOOTES  \citep{Kenter2005}.   These  programs  use
primarily  X-rays to  locate AGN  and multiwavelength  observations to
study their hosts. As  a result observational constraints to different
AGN  fueling models  have  become  available in  the  last few  years,
although the  results are  often contradictory and  the interpretation
may  vary  among  groups.   Morphological studies  for  example,  have
identified a  large fraction  of disks ($\approx  30$ per  cent) among
X-ray  selected  AGN at  $z\approx1$,  suggesting  that major  mergers
cannot  be   the  dominant   mode  of  SMBH   growth  at   that  epoch
\citep{Pierce2007, Gabor2009,  Georgakakis2009, Cisternas2011}.  X-ray
AGN  at $z\approx1$  are typically  found in  small and  moderate size
groups  \citep{Georgakakis2008_groups,  Silverman2009_env}  with  dark
matter    halo    masses    $\rm    \approx10^{12}-10^{13}\,M_{\odot}$
\citep{Coil2009, Hickox2009}.  This environment  is proposed to be the
most     conducive    to     merger     events    \citep[][but     see
  \citealt{Allevato2011}]{Hopkins2008_sam}.     Galaxy   interactions,
particularly those which include at least one early-type (red) system,
are  indeed  observed to  be  more  common  in group  environments  at
$z\approx1$  \citep{Lin2010}.  The star-formation  rate of  X-ray AGN,
although still not  well constrained, is suggested to  depend on their
accretion luminosity \citep[e.g.   ][]{Lutz2010, Shao2010}.  Below the
break  of  the  X-ray  luminosity  function it  is  claimed  that  the
star-formation  is  decoupled to  AGN  activity,  consistent with  the
stochastic  accretion mode  scenario.   Above the  knee  of the  X-ray
luminosity function it is  argued that the accretion luminosity scales
with  the  star-formation rate,  in  agreement  with the  major-merger
formation picture.

The observational  studies above essentially probe  active SMBHs close
to the peak of the accretion  power of the Universe, $z\approx 1$, and
lack  the  volume  to   provide  meaningful  constraints  on  the  AGN
population at low redshift $z\la 0.5$. This is an important gap in AGN
evolution studies, as there are  suggestions that the dominant mode of
SMBH accretion changes with  redshift, thereby leading to the observed
rapid     decline    of     the    accretion     power     at    $z<1$
\citep{Hopkins_Hernquist2006,         Hasinger2008,         Lutz2010}.
\cite{Fanidakis2011} for  example, propose  two main channels  of SMBH
growth, disk instabilities, which dominate at high redshift and bright
AGN luminosities,  and hot gas  accretion, which becomes  important at
low  redshift and  faint  luminosities.  In  the  models above  shifts
between  different  fueling  modes  with cosmic  time  should  imprint
detectable changes in the properties  of AGN hosts from $z=0$ to $z=1$
and beyond.


The  Sloan  Digital   Sky  Survey  \citep[SDSS;][]{Abazajian2009}  has
identified  the largest  sample  of low  redshift ($z\approx0.1$)  AGN
todate,      using      diagnostic      emission      line      ratios
\citep{Kauffmann2004}. The selection  function of that sample however,
is very  different from that of  X-ray AGN in  deep surveys, rendering
the comparison  difficult. Recent work has demonstrated  that there is
only  partial overlap  between AGN  samples selected  using diagnostic
emission lines and  X-rays \citep[e.g. ][]{Yan2011, Juneau2011}.  This
might  be due  to obscuration  effects, which  are likely  to  be more
severe  at X-rays,  dilution  of  the AGN  optical  emission lines  by
stellar  continuum  or  nebular   emission  lines  from  HII  regions,
intrinsic scatter in  the X-ray to optical luminosity  ratio of AGN or
the controversial  nature of  sources such as  LINERs.  This  class of
objects  are often  included  in AGN  samples  selected by  diagnostic
emission  line  ratios,  although  their  emission lines  may  not  be
dominated  by accretion  onto a  SMBH  \citep[e.g.][]{Sarzi2010}.  The
study  of  AGN hosts  as  a  function of  time  would  benefit from  a
homogeneous selection of active SMBHs at all redshifts.

A  step in  this  direction is  the  extended Chandra  Multiwavelength
Project \citep[ChaMP,][]{Green2009}, which detects serendipitous X-ray
sources on archival Chandra images. That survey covers an area of $\rm
32\,deg^2$ and therefore  probes a large enough volume  to detect in a
systematic way  X-ray AGN  at low redshift  \citep{Haggard2010}.  More
recently    \cite{Georgakakis_Nandra2011}   combined    archival   XMM
observations with  the SDSS (XMM/SDSS  survey) to compile a  sample of
X-ray selected  AGN at  low redshift, $z\approx0.1$,  over an  area of
$\rm 120\,deg^2$, much larger  than the extended ChaMP.  The advantage
of serendipitous X-ray surveys like the XMM/SDSS and the ChaMP is that
the AGN selection function at low redshift is almost identical to that
of  X-ray AGN  at  $z\approx1$  and beyond  in  deep surveys,  thereby
facilitating  the  comparison  by  minimising  differential  selection
effects.  \cite{Georgakakis_Nandra2011} demonstrated the power of this
approach.   By comparing X-ray  AGN in  the XMM/SDSS  at $z\approx0.1$
with those in the AEGIS  at $z\approx1$ they found little evolution of
the  fraction of  AGN in  red/blue hosts.   They argued  that  this is
evidence against a change of  the dominant SMBH fueling mechanism with
redshift.

In this  paper we expand on  the \cite{Georgakakis_Nandra2011} results
and provide new estimates of the volume density evolution of X-ray AGN
in red and blue hosts  by combining three different X-ray surveys with
median  redshifts  $z=0.1$, 0.3  and  0.8.  We  also determine  how  the
fraction of  X-ray AGN in  the overall galaxy population  changes with
redshift.  These results are discussed in the context of semi-analytic
models  for  the  growth  of  SMBHs  and  the  evolution  of  the  AGN
population.  Throughout this paper we adopt  $\rm H_{0} = 100 \, km \,
s^{-1} \, Mpc^{-1}$, $\rm  \Omega_{M} = 0.3$ and $\rm \Omega_{\Lambda}
=   0.7$.    Rest  frame   quantities   (e.g.   absolute   magnitudes,
luminosities, stellar masses) are parametrised by $h=H_{0} / 100$.

\section{X-ray AGN samples}\label{sec_xagn}

Three  different  datasets are  used  to  explore  the evolution  from
$z\approx0.1$ to $z\approx1$  of the space density of  AGN in blue and
red   cloud   hosts.    The   XMM/SDSS  serendipitous   X-ray   survey
\citep{Georgakakis_Nandra2011} is used to select AGN at $z\approx0.1$.
The  Needles in  the Haystack  Survey (NHS;  Georgantopoulos  2005) is
employed to study X-ray AGN at a median redshift of $z\approx0.3$.  At
$z\approx0.8$,  the AEGIS \citep{Davis2007},  the GOODS-North  and the
GOODS-South fields are combined to explore the properties of the X-ray
AGN population.  The number of X-ray AGN with $L_X (\rm 2 - 10 \, keV)
>10^{41}\,  erg \,  s^{-1}$  in each  of  the above  three samples  is
presented in Table \ref{tab_sample}.

\begin{figure}
\begin{center}
\includegraphics[height=0.9\columnwidth]{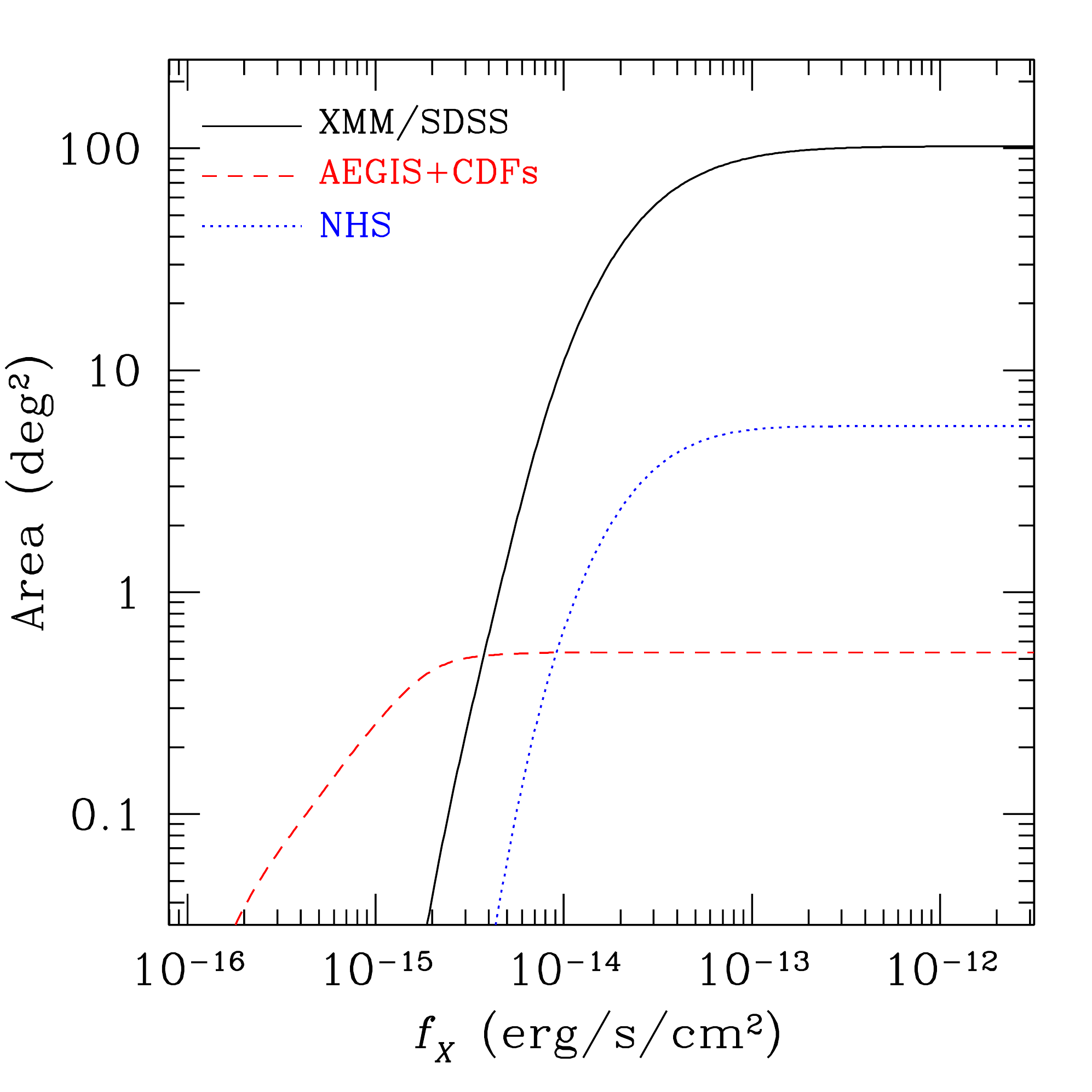}
\end{center}
\caption{Sensitivity curves for the XMM/SDSS survey (black solid line;
  2-10\,keV band), the NHS (blue  dotted line; 2-10\,keV band) and the
  combined AEGIS, GOODS-North and GOODS-South fields (red dashed line;
  0.5-10\,keV band).}\label{fig_curve}
\end{figure}

\subsection{The XMM/SDSS survey}

The XMM/SDSS  is a serendipitous  XMM survey of  the SDSS area  and is
designed    to    study   the    properties    of    X-ray   AGN    at
$z\approx0.1$. Details  on the reduction of the  XMM observations, the
detection of  sources, the estimation  of fluxes and  hardness ratios,
the determination of the  survey sensitivity and the identification of
X-ray   sources   with   optical   counterparts   are   described   in
\cite{Georgakakis_Nandra2011}.

The XMM/SDSS survey includes pointings which have clusters of galaxies
as their prime targets.
The overdensity of  sources in those fields may  bias estimates of the
fraction  of red/blue AGN  hosts or  the fraction  of AGN  relative to
galaxies.   We therefore  exclude  from the  analysis XMM/SDSS  survey
fields that have clusters  as their prime targets.  These observations
are identified from the target  name keyword of the event files.  This
reduces  the  XMM/SDSS  survey  area  to  102\,$\rm  deg^2$.   Sources
detected in the 2-8\,keV band  are used.  This energy interval is less
affected by obscuration biases  compared to softer energies, while the
XMM's sensitivity remains high, thereby  resulting in a sample that is
sufficiently large for statistical studies.  The hard band sensitivity
curve is plotted in Figure \ref{fig_curve}.

The low  redshift X-ray subsample  of the XMM/SDSS survey  consists of
235   hard-band   (2-8\,keV)    detections   with   $0.03<z<0.2$   and
$r<17.77$\,mag    after    correcting    for    Galactic    extinction
\citep{Schlegel1998}.  The  magnitude cut corresponds to  the limit of
the  SDSS Main  Galaxy  Sample \citep[$r<17.77$\,mag,][]{Strauss2002},
which  provides the  vast  majority  of redshifts  in  the SDSS.   The
photometry is from the  New York University Value-Added Galaxy Catalog
\citep[NYU-VAGC,][]{Blanton2005}  which uses  data from  the  SDSS DR7
\citep{Abazajian2009}.   The   NYU-VAGC  provides  better  photometric
calibration of  the SDSS data \citep{Padmanabhan2008}  compared to the
DR7 and an accurate description of the SDSS window function.

XMM prime targets are included in the XMM/SDSS source catalogue.  They
are identified by comparing the positions of the detected sources with
either the NED coordinates corresponding to the target name keyword of
the XMM  observations or  the right ascension  and declination  of the
nominal pointing of  the XMM observations.  A total  of 42 sources are
identified  as prime targets  in the  subsample with  $0.03<z<0.2$ and
$r<17.77$\,mag and are excluded from the analysis.  We retain however,
X-ray sources which lie at similar  redshifts as the prime target of a
particular X-ray observation.

X-ray fluxes are estimated in  the standard 2-10\,keV spectral band by
extrapolating  from the  observed count  rate in  the  2-8\,keV energy
interval (see  Georgakakis \& Nandra 2011 for  details).  The XMM/SDSS
survey detects  sources with  luminosities as low  as $L_X  \rm (2-10)
\approx 10^{40} \, erg \, s^{-1}$ (see section 4 for the estimation of
X-ray luminosity).  In contrast, the higher redshift samples (see next
sections) are sensitive  to $L_X \rm (2-10) \approx  10^{41} \, erg \,
s^{-1}$.  To make fair the comparison of AGN at different redshifts we
apply a luminosity cut of $L_X  \rm (2-10) = 10^{41} \, erg \, s^{-1}$
and  exclude sources  fainter  that this  limit.   The final  XMM/SDSS
sample used in  this paper consists of 175  serendipitous sources with
$L_X  \rm (2-10)  >  10^{41}  \, erg  \,  s^{-1}$, $r<17.77$\,mag  and
$0.03<z<0.2$ (median redshift $z=0.1$).

\subsection{The Needles in the Haystack Survey}

The original NHS is a  serendipitous XMM survey which was initiated to
select and study normal  galaxies, i.e.  non-AGN, at X-ray wavelengths
\citep{Georgantopoulos2005}.  That  survey consisted of  high Galactic
latitude ($b >  20$) XMM observations that became  public in June 2004
and overlapped with the second data release of the SDSS.  The original
NHS     is     now    superseded     by     the    XMM/SDSS     survey
\citep{Georgakakis_Nandra2011}.      However,     extensive    optical
spectroscopy has  been obtained for X-ray  sources in a  subset of the
original NHS fields (see below).   Therefore, we retain that survey in
its  original format  but  reanalyse the  XMM  observations using  the
pipeline  developed for the  reduction of  the XMM/SDSS  survey fields
\citep{Georgakakis_Nandra2011}.   As a  result  the source  detection,
flux  estimation, astrometric  corrections, optical  identification of
the X-ray sources and sensitivity  map construction in the updated NHS
follow the methodology described in \cite{Georgakakis_Nandra2011}.

\begin{figure}
\begin{center}
\includegraphics[height=0.9\columnwidth]{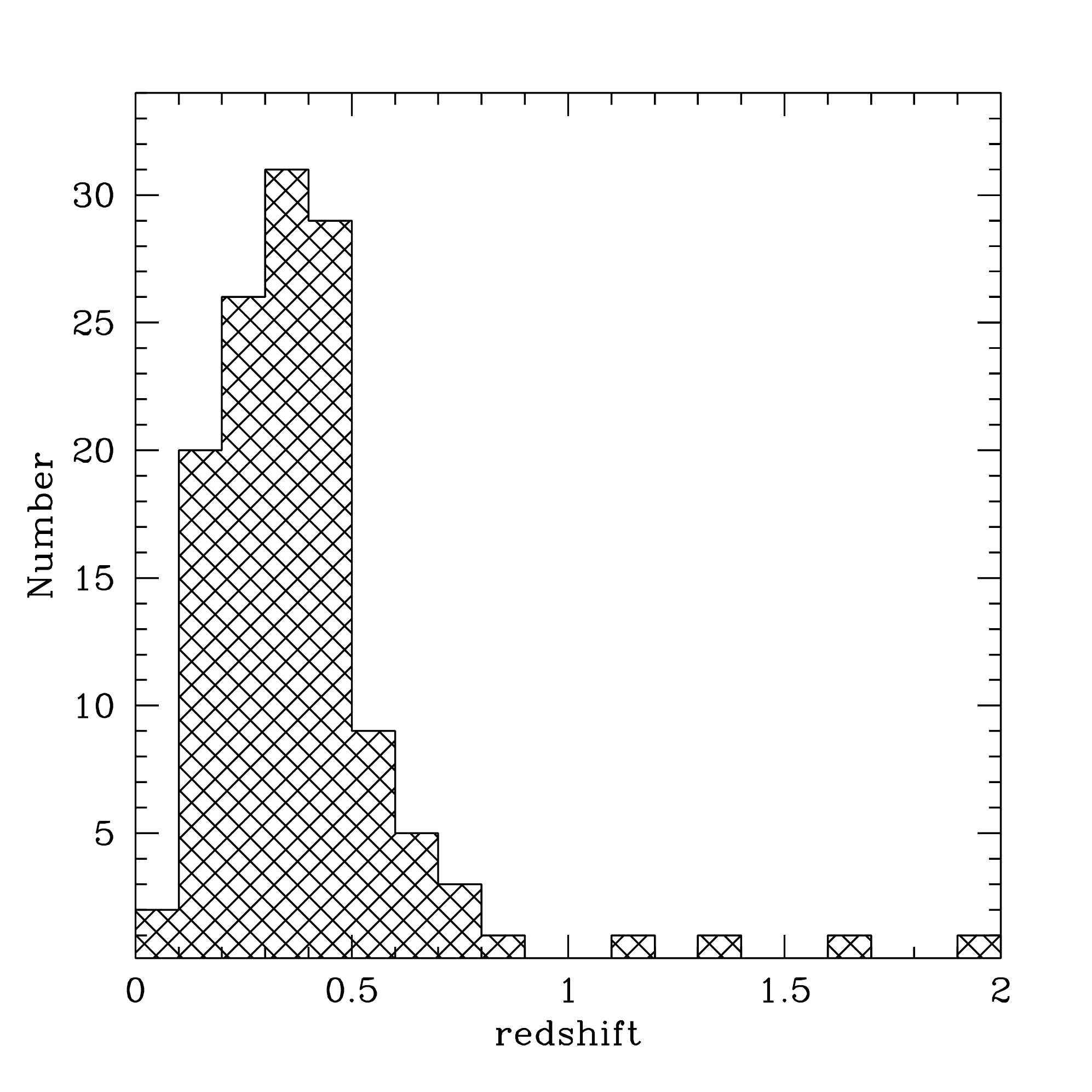}
\end{center}
\caption{Redshift distribution of NHS X-ray sources with spectroscopic
  follow-up  observations   from  our   ESO  VLT  and   NTT  programs.
}\label{fig_zdist}
\end{figure}

For  the   subset  of  the   original  NHS  fields   with  declination
$\delta<10$\,deg follow-up spectroscopy was carried out at the ESO VLT
and  NTT  telescopes\footnote{Based  on  observations  made  with  ESO
  Telescopes at the La Silla and Paranal Observatories under programme
  IDs 078.B-0623A and 080.B-0409A}.  The total area of that subsample,
which hereafter will be referred  to as NHS, is $\rm 5.6\,deg^2$.  All
X-ray  sources with  extended optical  light profiles  (i.e.  optically
resolved)   in  the   SDSS  and   $r<20.5$\,mag  have   been  observed
spectroscopically.  Because the  density of the targets on  the sky is
low,  each  source  was  observed  separately with  a  longslit.   The
observations were carried  out between October 2006 and  March 2008 in
queue  mode   with  the  FORS2  (FOcal  Reducer   and  low  dispersion
Spectrograph) on  the VLT (Very  Large Telescope) and the  EMMI (ESO's
Multimode Instrument) at the NTT (New Technology Telescope).  The data
were  reduced  using  standard   routines  in  IRAF.   Redshifts  were
determined for 129 out of 151 targeted sources by visual inspection of
the  reduced spectra.   The  redshift measurements  correspond to  at
least two  secure spectral features.   Figure \ref{fig_zdist} presents
the redshift distribution  of the sample and demonstrates  that at the
magnitude  limit $r=20.5$\,mag  the  majority of  the  sources lie  at
$z\la0.5$.

In  the  updated NHS  catalogue  there  are  126 hard-band  (2-8\,keV)
detected  sources  with   $r<20.5$\,mag  and  extended  optical  light
profiles.  Redshift estimates are  available for 100 of them (including
one  Galactic star).  The  redshift incompleteness  is because  of (i)
featureless spectra, (ii)  failure to observe some of  the targets and
(iii)  mismatches between  the updated  NHS source  catalogue  and the
original  one (Georgantopoulos  et al.   2005) used  to  construct the
target list for follow-up spectroscopy.  The sample used in this paper
consists of 73 hard X-ray sources with $0.1<z<0.5$, $r<20.5$\,mag. The
median redshift  is $z\approx0.3$. The  sensitivity curve of the  NHS in
the hard band is plotted in Figure \ref{fig_curve}.

Optically  unresolved X-ray  sources are  underrepresented in  the NHS
spectroscopic  sample  as they  have  not  been  targeted by  our  ESO
observations.  The majority of  those sources however, are expected to
lie at redshift $z>0.5$. Out  of the 124 unresolved X-ray sources with
$r<20.5$\,mag, a total of 28 have redshift available from the SDSS. Of
them only one lies in the redshift interval $0.1<z<0.5$.  We therefore
expect that the exclusion of optically unresolved X-ray sources is not
going to  have a  strong effect on  the results and  conclusions. This
issue is further discussed in the results section.

\subsection{The deep X-ray surveys}

AGN at $z\approx0.8$ are selected from the Chandra surveys of the AEGIS,
GOODS-North (2\,Ms  Chandra Deep  Field North) and  GOODS-South (2\,Ms
Chandra  Deep Field  South).  The  Chandra observations  in  the those
fields have been  reduced and analysed in a  homogeneous way using the
methodology described by \cite{Laird2009}. The optical counterparts of
the  X-ray sources  have been  identified using  the  Likelihood Ratio
method as described in  \cite{Georgakakis2009}. For the AEGIS field we
use the $BRI$ ground based  optical photometry obtained as part of the
DEEP2 spectroscopic  survey \citep{Coil2004}. For the  GOODS North and
South fields the optical identification  of the X-ray sources uses the
Hubble Space Telescope (HST) Advanced Camera for Surveys (ACS) imaging
observations     of     those     fields     \citep[see     ][     for
  details]{Georgakakis2009}.

The spectroscopy for  X-ray sources in the AEGIS is  from a variety of
sources:  the  DEEP2   redshift  survey  \citep{Davis2003},  follow-up
observations of  X-ray sources  at the MMT  using the  Hectospec fiber
spectrograph \citep{Coil2009}, the SDSS  and a number of spectroscopic
programs  that   targeted  galaxies   in  the  original   Groth  Strip
\citep{Weiner2005}.   Optical  spectroscopy   in  the  GOODS-North  is
available from either programmes  that specifically targeted the X-ray
population  in  these  fields  \citep[e.g.][]{Barger2003,  Barger2005,
  Cowie2003}     or    the     Keck    Treasury     Redshift    Survey
\citep[TKRS;][]{Wirth2004}.  There are many spectroscopic campaigns in
the CDF-South. These include follow-up observations targeting specific
populations,  such  as  X-ray  sources  \citep{Szokoly2004},  $K$-band
selected  galaxies \citep{Mignoli2005},  and high  redshift candidates
\citep[e.g.][]{Dickinson2004, Stanway2004a,  Stanway2004b}, as well as
generic  spectroscopic surveys  of faint  galaxies in  the  GOODS area
using   the  FORS2   \citep{Vanzella2005,   Vanzella2008}  and   VIMOS
\citep{LeFevre2004,     Popesso2008,     Popesso2009,    Balestra2010}
spectrographs at the VLT.

We select X-ray  sources detected in the 0.5-7\,keV  band with optical
magnitudes  $R_{AB}<24.1$\,mag for  the  AEGIS and  $i_{AB}<23.5$\,mag
(F775W HST/ACS  band) for the  GOODS-North and South fields.   At this
magnitude  cut the  spectroscopic  identification rate  of the  AEGIS,
GOODS-North  and  GOODS-South  X-ray   sources  is  67  (375/557),  85
(118/139)  and   87  (105/121)  per  cent,   respectively.   From  the
spectroscopically identified sources we select those with redshifts in
the interval $0.6<z<1.2$.   This results in 149 sources  in the AEGIS,
66 in the GOODS-North and  54 in the GOODS-South.  The median redshift
of the sample is 0.8.  The  choice of the X-ray band for the detection
of sources is  to ensure that our $z\approx0.8$  sample is selected at
similar   rest-frame  energies   (1-14\,keV)   as  the   $z\approx0.1$
(2-9\,keV)   and  $z\approx0.3$  samples   (3-10\,keV).   Differential
selection  effects  are  therefore  small,  thereby  facilitating  the
comparison  of  the  AGN  properties  among the  three  samples.   The
combined X-ray sensitivity  curve for the 3 fields  is shown in Figure
\ref{fig_curve}.

\section{Galaxy Samples}

In the next  sections the space density of X-ray  AGN will be compared
with  that of  the overall  galaxy population.   We  therefore compile
three spectroscopic galaxy samples at $z\approx0.1$, $z\approx0.3$ and
$z\approx0.8$,  which  probe similar  redshift  intervals  as the  low
redshift subset  of the XMM/SDSS  survey, the NHS and  the AEGIS+GOODS
X-ray AGN samples respectively. The  number of galaxies in each sample
is listed in Table \ref{tab_sample}.

At  $z\approx0.1$, we  select a  total of  9541 NYU-VAGC  sources with
$r<17.77$\,mag  that overlap  with the  XMM/SDSS survey  fields (after
excluding galaxy  cluster pointings) and  have spectroscopic redshifts
in the interval $0.03<z<0.2$.

The  data release 3  of the  DEEP2 spectroscopic  survey in  the AEGIS
field is used to select galaxies at $z\approx0.8$.  This spectroscopic
program uses  the DEIMOS  spectrograph \citep{Faber2003} on  the 10\,m
Keck-II  telescope  to obtain  redshifts  for  galaxies  to $R_{AB}  =
24.1$\,mag.  The observational setup uses a moderately high resolution
grating ($R\approx5000$),  which provides a velocity  accuracy of $\rm
30\,km\,s^{-1}$  and a wavelength  coverage of  6500--9100\,\AA.  This
spectral  window  allows  the  identification of  the  strong  [O\,II]
doublet  3727\AA\, emission line  to $z<1.4$.   We use  DEEP2 galaxies
with redshift  determinations secure  at the $>90\%$  confidence level
(quality flag $Q \ge 3$; Davis  et al. 2007). A total of 6797 galaxies
with redshifts $0.6<z<1.2$ are selected to determine the space density
of the galaxy population at  a median redshift of $z\approx0.8$ and to
compare against that of X-ray AGN.
 
Although the  AEGIS DEEP2 spectroscopic  survey is geared  toward high
redshift  sources, it  also includes  a  large number  of galaxies  at
$z<0.6$ (4601).   We select a total  of 1960 galaxies  in the interval
$0.2<z<0.4$ to estimate their luminosity function at $z\approx0.3$ and
compare it with that of X-ray AGN identified in the NHS.

\begin{table}
\caption{X-ray AGN and galaxy samples}\label{tab_sample}
\begin{center} 
\scriptsize
\begin{tabular}{l c c  c c c}

\hline  \multicolumn{6}{c}{X-ray  AGN}\\ \hline  sample  & redshift  &
number & red & blue & AGN \\ & & & hosts & hosts & dominated \\ \hline

 XMM/SDSS & 0.03-0.2 & 175 & 76 & 68 & 31 \\

 NHS & 0.1-0.5 & 73 & 24 & 27 & 22 \\

 AEGIS/GOODS & 0.6-1.2 & 269 & 124 & 111 & 34 \\

 AEGIS/GOODS (stellar mass) & 0.6-1.2 & 217 & 98 & 97 & 22 \\

\\\hline

\multicolumn{6}{c}{Galaxies}\\ \hline sample & redshift & number & red
& blue & -- \\ & & & hosts & hosts & \\ \hline

 XMM/SDSS & 0.03-0.2 & 8643 & 5304 & 3339 & -- \\

 AEGIS-NHS & 0.2-0.4 & 1960 & 291 & 1669 & -- \\

 AEGIS & 0.6-1.2 & 6797 & 1383 & 5414 & -- \\

AEGIS (stellar mass) & 0.6-1.2 & 3436 & 648 & 2788 & -- \\\hline

\end{tabular} 
\begin{list}{}{}
\item 
The columns are: (1): X-ray AGN or galaxy sample; (2): redshift range of the
sources in each sample; (3) total number of sources in the sample; (4)
number of  red cloud sources ;  (5) number of blue  cloud sources; (6)
AGN with broad  lines and/or and dominant nuclear  point source in the
optical.
\end{list}
\end{center}
\end{table}

\section{Rest-frame properties}

This section describes how  the rest-frame colour, stellar mass, X-ray
luminosity and intrinsic column density  of galaxies and X-ray AGN are
determined.

The    {\sc   kcorrect}    version   4.2    routines    developed   by
\cite{Blanton_Roweis2007}  are  used to  fit  spectral  models to  the
optical  photometry of X-ray  sources and  galaxies and  then estimate
rest-frame colours and  absolute magnitudes in the AB  system. For the
sake  of uniformity,  in  all three  samples  used in  this paper  we
estimate  rest-frame  magnitudes in  the  $^{0.1}u$, $^{0.1}g$  bands,
which are the SDSS $u$, $g$ filters shifted to $z=0.1$.  The advantage
of this choice of bandpasses is twofold.  They are close to observed
frame 
for  our  $z\approx0.1$  XMM/SDSS   sample  and  also  have  effective
wavelengths that  are similar to the  rest-frame effective wavelengths
of the DEEP2 $R$ and $I$ filters at $z=1$ \citep{Blanton2006}.

For galaxies selected in the SDSS and X-ray AGN in either the XMM/SDSS
or the  NHS the  {\sc kcorrect} routines  are applied to  the observed
$ugriz$ SDSS photometry.   The rest-frame $^{0.1}u$ and $^{0.1}g$-band
magnitudes are estimated from the apparent magnitudes of the source in
the $u$ and $g$ band filters respectively. In the AEGIS field the {\sc
  kcorrect} routines  are used to  fit models to the  $BRI$ photometry
obtained  as  part  of  the  DEEP2 survey  \citep{Coil2004}  and  then
estimate the  rest-frame $^{0.1}u$ and  $^{0.1}g$-band magnitudes from
the  observed  DEEP2 $R$  and  $I$-band  data  respectively.  For  the
GOODS-North  and  South  fields  the  HST/ACS photometry  in  the  $B$
(F435W), $V$ (F606W), $i$ (F775W)  and $z$ (F850LP) bandpasses is used
as input to {\sc kcorrect}. The rest-frame magnitudes in the $^{0.1}u$
and $^{0.1}g$ bands  are determined from the F435W  and F775W apparent
magnitudes    respectively.    The   $^{0.1}(u-g)$    vs   $M_{^{0.1}g}$
colour-magnitude diagrams of AGN and  galaxies at z=0.1 and z=0.8 have
been  presented  by \cite{Georgakakis_Nandra2011}.   In  that study  a
colour cut of  $^{0.1}(u-g)=1.4$ was employed to separate  blue from red
cloud galaxies.  This value is also used here to define AGN subsamples
associated with red/blue hosts.

For the estimation of stellar  mass, $M_{star}$, we adopt the relation
between $B$-band mass--to--light ratio  and rest-frame $B-R$ colour of
\cite{zibetti2009}.   We  choose  to  use  the  mass--to--light  ratio
coefficients estimated for the  Jonhson-Cousin filterset (see Table B1
of Zibetti et al. 2009) instead  of the SDSS bands.  It was found that
the latter set of coefficients, when applied to the AEGIS/DEEP2 galaxy
sample, produced  too many massive galaxies  at $z\approx0.8$ compared
to the mass function of Borch et al.  (2006, see results section). The
DEEP2 $BRI$ photometric bands are  better suited for the estimation of
the  rest-frame $U-B$  colour of  galaxies at  $z\approx1$,  for which
\cite{zibetti2009} do not  provide mass--to--light ratio coefficients.
Therefore for the estimation of  stellar masses for AEGIS galaxies and
AGN  we  use the  $ugriz$  photometry obtained  as  part  of the  deep
synoptic  Canada-France-Hawaii  Telescope  Legacy Survey  (CFHTLS)  to
determine their  $B-R$ colours and $B$-band luminosities.   We use the
CFHTLS            data            from            the            T0003
release\footnote{http://terapix.iap.fr/article.php?idarticle=556.},
which      reaches     a      limiting      magnitude     of      $\rm
i_{AB}\approx26.5\,mag$. The overlap between the CFHTLS photometry and
the DEEP2 spectroscopic survey  is $\rm 0.3\,deg^2$.  The total number
of spectroscopically identified galaxies  and X-ray AGN in the overlap
region are listed in Table \ref{tab_sample}.  The adopted approach for
determining  stellar  masses is  not  as  accurate  as fitting  galaxy
templates to the  observed spectral energy distribution. Nevertheless,
as it  will be shown  in the next  sections, it provides  stellar mass
estimates which are  adequate for our purposes. Also,  the results and
conclusions are not sensitive  to systematics (i.e.  choise of initial
mass function)  which affect absolute stellar mass  estimates. This is
because the stellar mass distribution of AGN hosts is studied relative
to that  of galaxies  (see Results section).   The calculation  of the
$B-R$ colour  and $B$-band absolute  magnitudes of galaxies  and X-ray
AGN is carried out using the {\sc kcorrect} routines.

The  intrinsic  column density,  $N_H$,  of  individual  X-ray AGN  is
determined from the hardness  ratios between the soft (0.5-2\,keV) and
the hard (2-7\,keV for AEGIS, GOODS, 2-8\,keV for XMM/SDSS, NHS) X-ray
bands  assuming  an  intrinsic  power-law X-ray  spectrum  with  index
$\Gamma=1.9$ \citep[e.g.][]{Nandra1994}.  The derived column densities
are  then used  to  convert the  X-ray  flux in  either the  2-10\,keV
(XMM/SDSS, NHS) or the  0.5-10\,keV (AEGIS, GOODS) bands to unabsorbed
X-ray luminosity in the rest-frame 2-10\,keV energy interval.

The  NHS and  AEGIS/GOODS X-ray  samples used  in this  paper  are not
sensitive to sources fainter than $L_X  \rm (2 - 10\,keV) = 10^{41} \,
erg  \,  s^{-1}$.   To   avoid  luminosity  dependent  biases  in  the
comparison  of the  properties X-ray  AGN host  galaxies  at different
redshifts we choose to use in  the analysis only sources with $L_X \rm
(2 - 10\,keV) > 10^{41} \,  erg \, s^{-1}$. This criterion excludes 18
XMM/SDSS  sources.   In many  studies  a  brighter  luminosity cut  is
adopted,  $\rm  >  10^{42}  \,  erg \,  s^{-1}$,  to  avoid  potential
contamination  of AGN  samples by  normal galaxies.   As  discussed by
Georgakakis \& Nandra (2011),  although normal galaxy {\em candidates}
are  present at  faint  X-ray luminosities,  AGN  remain the  dominant
population.  We provide  an estimate of the level  of contamination by
normal  galaxies by  calculating the  fraction of  X-ray  sources with
X-ray--to--optical  flux ratio  $\log f_X/f_{opt}<-2$  and  soft X-ray
spectra, $\rm  N_H<10^{22} \, cm^{-2}$.  These two criteria  are often
adopted  in the  literature  for selecting  normal  galaxies in  X-ray
surveys    \citep[e.g][]{Hornschemeier2003,   Georgakakis2006}.    The
fraction of X-ray sources fulfilling the above criteria are 3 per cent
(5/175) in the XMM/SDSS and 1  per cent (3/269) for the combined AEGIS
and CDF fields. There are no sources with $\log f_X/f_{opt}<-2$ in the
NHS.  Contamination by non-AGN is therefore not expected to affect our
results and conclusions.

\section{Estimation of the space density of AGN and galaxies}\label{sec_xlf}

The space density  of X-ray AGN and galaxies in  bins of stellar mass,
  (mass-function;   MF),  $M_{^{0.1}g}$  absolute  magnitude
(optical  luminosity function;  OLF) and  2-10\,keV  luminosity (X-ray
luminosity function; XLF) is derived using the standard non-parametric
$\rm  1/V_{max}$ method \citep{Schmidt1968}.   In this  calculation we
take into  account the  X-ray selection function  (in the case  of the
AGN),  the  optical  magnitude  limit  of  different  samples  and  the
spectroscopic incompleteness.  The MF, OLF and XLF in logarithmic bins
are estimated by the relation

\begin{equation} \phi_Y \, dY = \sum_{i} \frac{w_{i}}{V_{max,i}},
\end{equation}

\noindent where Y equals $\log  M_{star}$ for the MF, $M_{^{0.1}g}$ in
the case  of the  OLF and $\log  L_X(\rm 2  - 10 \,  keV)$ in  the XLF
calculation. In  the equation above  $w_{i}$ is the weight  applied to
each  spectroscopically  identified  source  $i$ to  correct  for  the
spectroscopic  incompleteness of  the different  samples  (see below).
$V_{max,i}$ is  the maximum comoving  volume for which the  source $i$
satisfies  the  sample   selection  criteria,  i.e.   redshift  range,
apparent optical  magnitude limit and,  in the case of  X-ray sources,
X-ray  flux limit.   For X-ray  AGN the  $V_{max,i}$ depends  on X-ray
luminosity   ($L_X$),  absolute  optical   magnitude  ($M_{^{0.1}g}$),
redshift ($z$) as  well as the overall shape of  the optical and X-ray
Spectral Energy Distributions (SED)

\begin{equation}\label{eq_vmax_x}
 V_{max,i}(L_X,M,z) =  \frac{c}{H_0} \int_{z1}^{z2} \, \Omega(L_X,z)\,
 \frac{dV}{dz}\,dz\, dL,
\end{equation}

\noindent where  $dV/dz$ is the  volume element per  redshift interval
$dz$.   The integration limits  are $z1=z_L$  and $z2=min(z_{optical},
z_U)$, where we have defined $z_L$, $z_U$ the lower and upper redshift
limits of  the sample and $z_{optical}$  is the redshift  at which the
source will  become fainter than  the survey optical  magnitude limit.
$\Omega(L_X,z)$ is the solid angle  of the X-ray survey available to a
source with  luminosity $L_X$ and  column density $N_H$ at  a redshift
$z$  (corresponding to  a flux  $f_X$ in  the X-ray  area  curve). For
galaxies  the $V_{max,i}$  is  a function  of  $M_{^{0.1}g}$, $z$  and
optical SED only and hence equation \ref{eq_vmax_x} simplifies to

\begin{equation}\label{eq_vmax_o}
V_{max,i}(M,z)   =   \frac{c}{H_0}   \,   \Omega   \int_{z1}^{z2}   \,
\frac{dV}{dz}\,dz\, dL,
\end{equation}

\noindent   where  the   symbols   are  the   same   as  in   equation
\ref{eq_vmax_x}  and   $\Omega$  is  the  survey   solid  angle.   The
uncertainty at a given luminosity or mass bin is

\begin{equation} 
\delta \phi_Y^2 = \sum_{i} \left ( \frac{w_{i}}{V_{max,i}} \right )^2.
\end{equation}

\noindent 
Sample variance is  not included in the error  budget. This is because
we  are mainly  interested in  the space  density of  AGN  relative to
galaxies selected  in the same  field. The effects of  sample variance
are therefore  minimised in this  differential approach.  This  is not
the case for the $z=0.3$ sample,  for which the X-ray AGN and galaxies
are selected  from different  fields.  Nevertheless, the  next section
shows that  the AGN XLFs and  the galaxy MFs/OLFs  determined from the
samples presented in this paper  at median redshifts z=0.1, 0.3 and 0.8,
are in  good agreement with previous estimates.  This further suggests
that sample variance effects are small.

The conversion  of the absolute  to apparent optical magnitude  in the
1/Vmax calculation  uses the  optical k-corrections determined  by the
{\sc  kcorrect} version  4.2 routines  \citep{Blanton_Roweis2007}. The
model that best fits the optical  photometric data of a source is also
used  to  estimate k-corrections  for  the  same  source at  different
redshifts. In the  case of the XLF, the  intrinsic $N_H$ of individual
X-ray  sources is  taken into  account in  the 1/Vmax  estimation. The
X-ray k-corrections  are calculated by adopting  an absorbed power-law
spectral  energy  distribution  with  $\Gamma=1.9$  and  photoelectric
absorption  cross  sections as  described  by \cite{Morrison1983}  for
solar metallicity.

For SDSS galaxies and AGN in  the XMM/SDSS survey, the weight $w_i$ in
equation 1,  which corrects  for spectroscopic incompleteness,  is the
fraction  of the SDSS  Main Galaxy  sample spectroscopic  targets that
have reliable redshifts in a given sector ({\sc fgotmain} parameter in
NYU-VAGC).   For the  NHS, AEGIS  and GOODS  X-ray sources  the weight
$w_i$  is estimated  following the  methodology of  \cite{Lin1999} and
\cite{Willmer2006}.    For  each   X-ray  source   $i$   targeted  for
spectroscopy the  number of X-ray sources with  (i) reliable redshifts
in a  given redshift  interval, $z_L<z<z_H$, (ii)  successful redshift
determinations outside that interval, $z<z_L$ or $z>z_H$, (iii) failed
redshifts and (iv) unavailable spectroscopic observations, are counted
within     spheres     defined     in    the     three     dimensional
colour-colour-magnitude space.  When  estimating the probability, $P$,
that a spectroscopically unidentified  source lies within the redshift
interval  $z_L<z<z_H$ it  is assumed  that unsuccessful  redshifts lie
outside   the  redshift   range  of   interest,  $z<z_L$   or  $z>z_H$
\cite[``minimal''  model of  ][]{Willmer2006}.  The  weight  $w_i$ for
each spectroscopic source  is the sum of the  probabilities $P$ of all
sources  within the colour-colour-magnitude  sphere. For  the redshift
limits ($z_L$,  $z_H$) we adopt the  values (0.1, 0.5)  for NHS, (0.2,
1.4)  for AEGIS  and  (0.6,  1.2) for  GOODS,  respectively. The  data
spheres  are defined  by the  $g-r$,  $r-i$ colours  and the  $r$-band
magnitude  in the  NHS,  the  $B-R$, $R-I$  colours  and the  $R$-band
magnitude in the AEGIS and  the $B-V$, $V-i$ and $i$-band magnitude in
the GOODS.  When estimating the  OLF of the overall  galaxy population
using data from the AEGIS DEEP2 spectroscopic survey we include in the
weights an  additional correction  which accounts for  the probability
that    a    galaxy    will    be    placed   on    a    DEEP2    mask
\citep[see][]{Willmer2006}.

\begin{figure}
\begin{center}
\includegraphics[height=0.9\columnwidth]{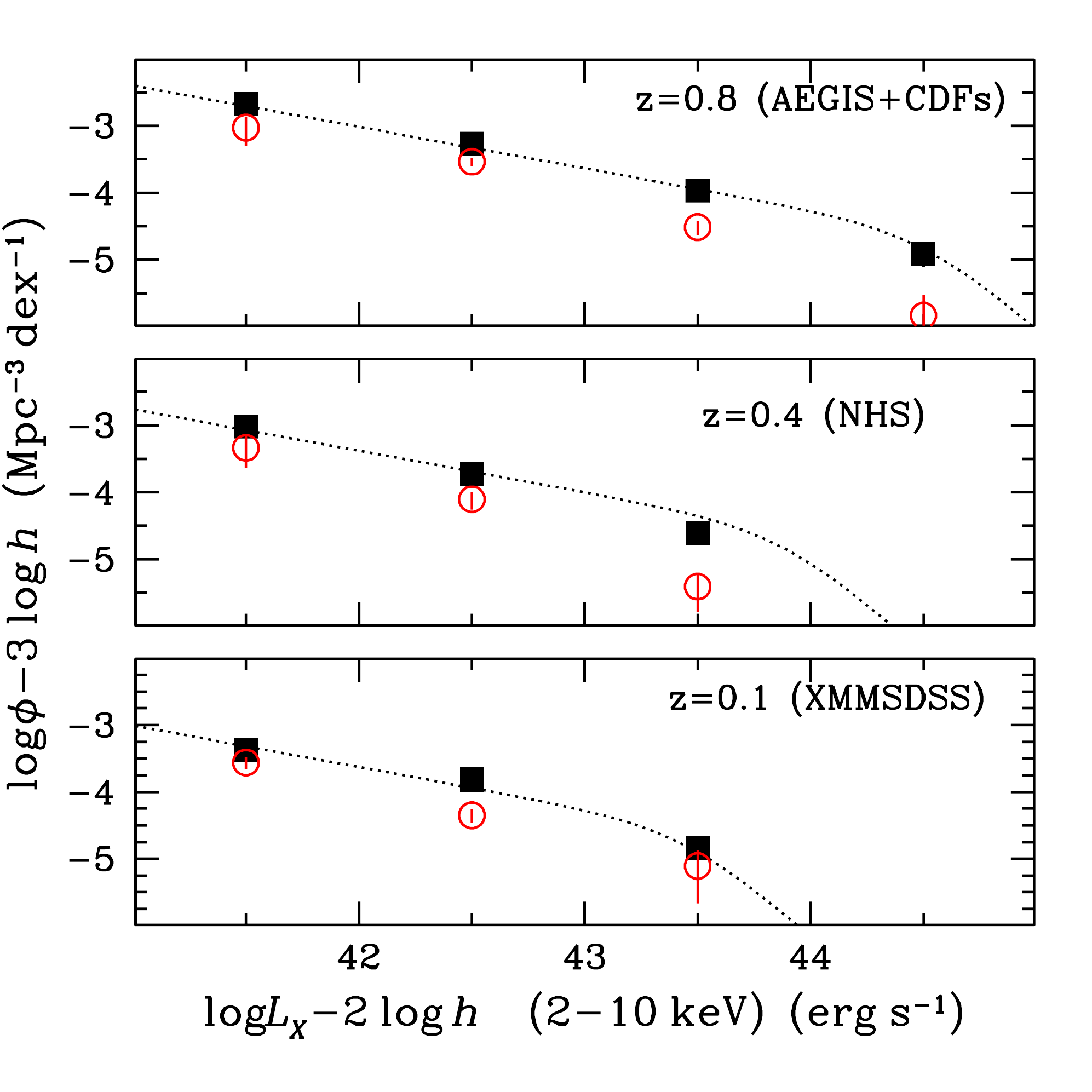}
\end{center}
\caption{The  2-10\,keV X-ray luminosity  function. The  black filled squares
  are  the XLF  estimates from  the XMM/SDSS  survey  at $z\approx0.1$
  (bottom  panel), the  NHS at  $z\approx0.3$ (middle  panel)  and the
  AEGIS+GOODS surveys  at $z\approx0.8$ (top panel).  The dotted lines
  corresponds to the Luminosity  And Density Evolution (LADE) model of
  Aird et al. (2010) estimated at  the median redshift of each sample. In
  all  panels  the red open circles  are AGN  in  red  host galaxies  with
  rest-frame colour $^{0.1}(u-g)>1.4$.  }\label{fig_xlf}
\end{figure}

\begin{figure}
\begin{center}
\includegraphics[height=0.9\columnwidth]{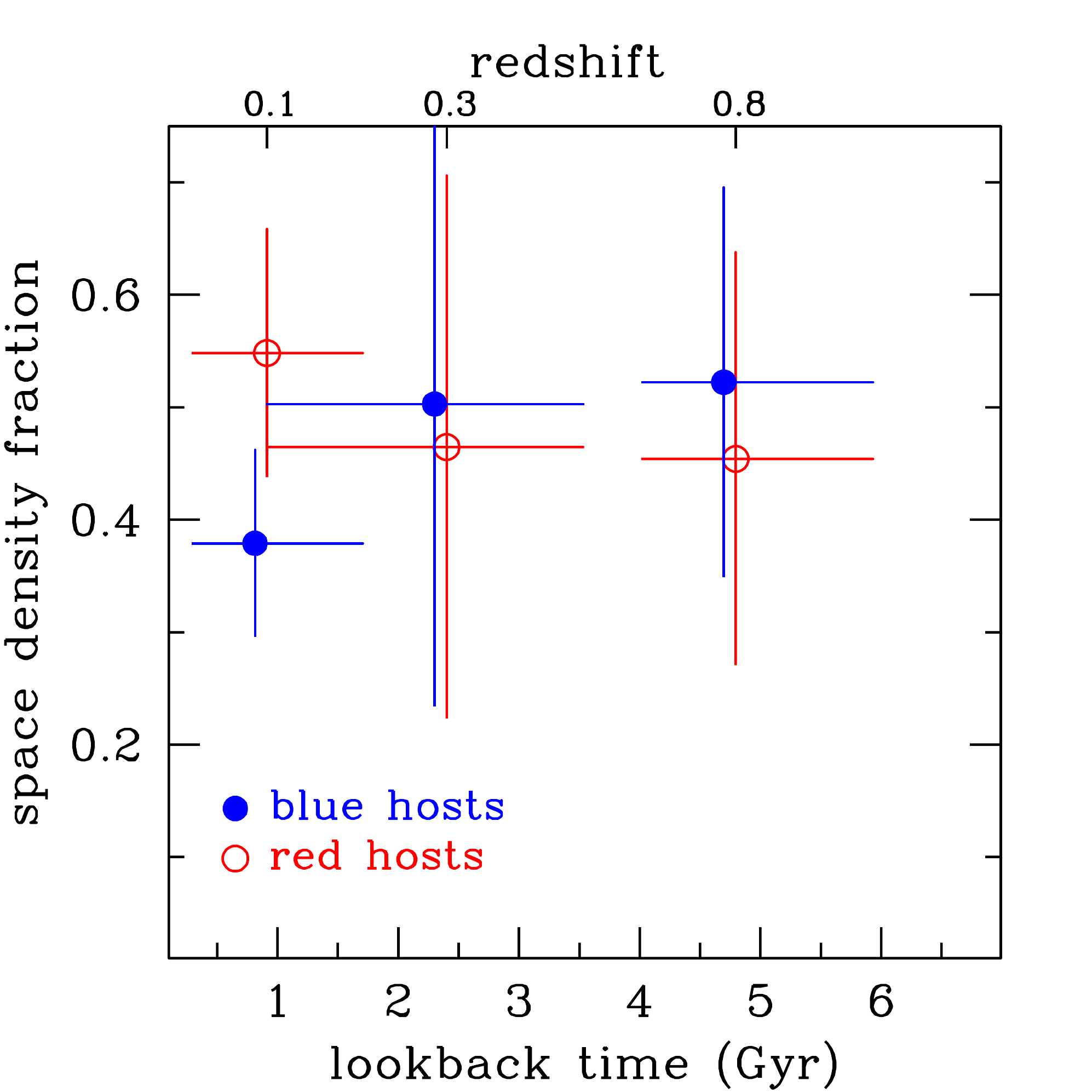}
\end{center}
\caption{Space density  fraction of X-ray  AGN in blue/red hosts  as a
  function of lookback time (lower-axis) and redshift (upper x-axis). The vertical
  axis  is  the  ratio  of  the 2-10\,keV  X-ray  luminosity  function
  (integrated  over all  luminosities  above $\rm  10^{41}  \, erg  \,
  s^{-1}$) of red  (filled circles) and blue (open  circles) AGN hosts
  divided by the integrated XLF  for all AGN.  The blue filled circles
  are  offset  by  -0.2\,Gyrs  for  clarity. The  errors  are  Poisson
  estimates  and are  propagated from  the uncertainties  of  the XLF.
}\label{fig_xlf_fraction}
\end{figure}

\begin{figure}
\begin{center}
\includegraphics[height=0.9\columnwidth]{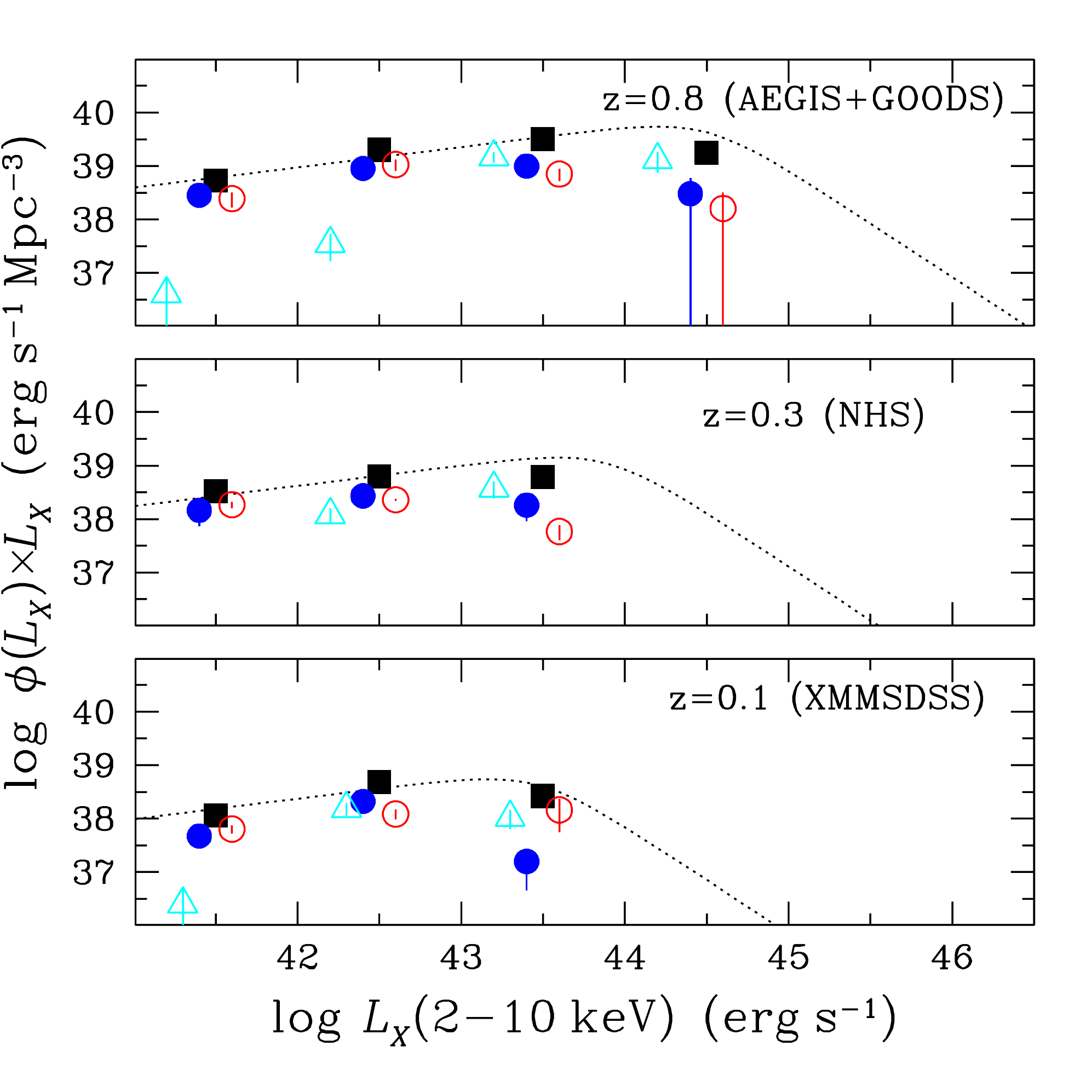}
\end{center}
\caption{X-ray  luminosity   density  as  a  function   of  hard  band
  luminosity.   The  black filled squares  are  for  the  XMM/SDSS survey  at
  $z\approx0.1$  (bottom  panel),  the  NHS at  $z\approx0.3$  (middle
  panel) and the AEGIS+GOODS surveys at $z\approx0.8$ (top panel). The
  dotted lines  corresponds to  the LADE model  of Aird et  al. (2010)
  estimated at the  median redshift of each sample.  In all panels the
  red circles are for AGN in  red host galaxies (offset by 0.1\,dex in
  luminosity for clarity), the filled  blue circles are for blue hosts
  (offset by -0.1\,dex  in luminosity) and the cyan  triangles are for
  X-ray AGN whose optical colours  are dominated by the central engine
  (offset  by  -0.2\,dex  in   luminosity).  The  errors  are  Poisson
  estimates.  }\label{fig_ld}
\end{figure}

\section{Results}

This section presents  the XLF, OLF and MF of X-ray  AGN, both for the
full XMM/SDSS, NHS, AEGIS+CDFs samples and for subsamples split by the
rest-frame colour of the host galaxy. In this exercise we identify and
exclude from the  analysis sources with either broad  emission lines in
their optical spectra and/or a prominent nuclear point source in their
optical  images (SDSS  or  HST/ACS survey  of  the AEGIS/GOODS).   The
optical continua of those sources  are contaminated by AGN light. This
introduces  biases in the  determination of  the host  galaxy absolute
optical magnitude and stellar mass. Also, the colours of those sources
do not provide information on the dominant stellar population of their
hosts.  We note however, that AGN dominated sources {\it are} included
in the determination of both the total XLF (i.e.  not split by colour)
and the total X-ray luminosity density.

\subsection{The X-ray luminosity function}

Figure  \ref{fig_xlf} and Table  \ref{tab_XLF} present  the 2-10\,keV
XLFs of the  XMM/SDSS, NHS and AEGIS+GOODS surveys  at median redshifts
of  0.1,  0.3  and  0.8  respectively. At  all  three  redshifts,  the
estimated XLFs are  in good agreement with the  Luminosity And Density
Evolution (LADE) model  of Aird et al.  (2010),  which is also plotted
in  the figure.   Although our  X-ray samples  are  apparent magnitude
limited,  this selection  effect  is correctly  accounted  for in  the
1/Vmax  calculation.  For  the  NHS in  particular,  the exclusion  of
optically unresolved X-ray sources, for which optical spectra have not
been  obtained  as  part   of  our  ESO  follow-up  programs  (section
\ref{sec_xagn}), does  not appear  to introduce incompleteness  in XLF
estimation.   This  is  consistent  with our  finding  that  optically
unresolved  sources are dominated  by AGN  outside the  redshift range
0.1-0.5 probed by the NHS.

Next we split the three  samples by rest-frame $^{0.1}(u-g)$ colour to
explore changes  with redshift  of the space  density of X-ray  AGN in
red/blue hosts and get insights  into the dominant accretion mode onto
SMBHs at  different epochs.  AGN  dominated sources are  excluded from
the  analysis.  The  total numbers  of red  cloud [$^{0.1}(u-g)>1.4$],
blue  cloud  [$^{0.1}(u-g)<1.4$]  and  AGN-dominated sources  in  each
sample are  listed in Table  \ref{tab_sample}.  The XLFs of  red cloud
AGN  are  shown  in  Figure  \ref{fig_xlf} and  are  listed  in  Table
\ref{tab_XLF}.  The  ratio of the integrated  XLF of AGN  in red hosts
over    the   total    integrated   XLF    is   plotted    in   Figure
\ref{fig_xlf_fraction}.  It demonstrates  that within the errors there
is no strong evidence for evolution of the red host AGN fraction since
$z\approx0.8$.   Similar results  are  obtained for  the  AGN in  blue
hosts.

\begin{figure}
\begin{center}
\includegraphics[height=0.9\columnwidth]{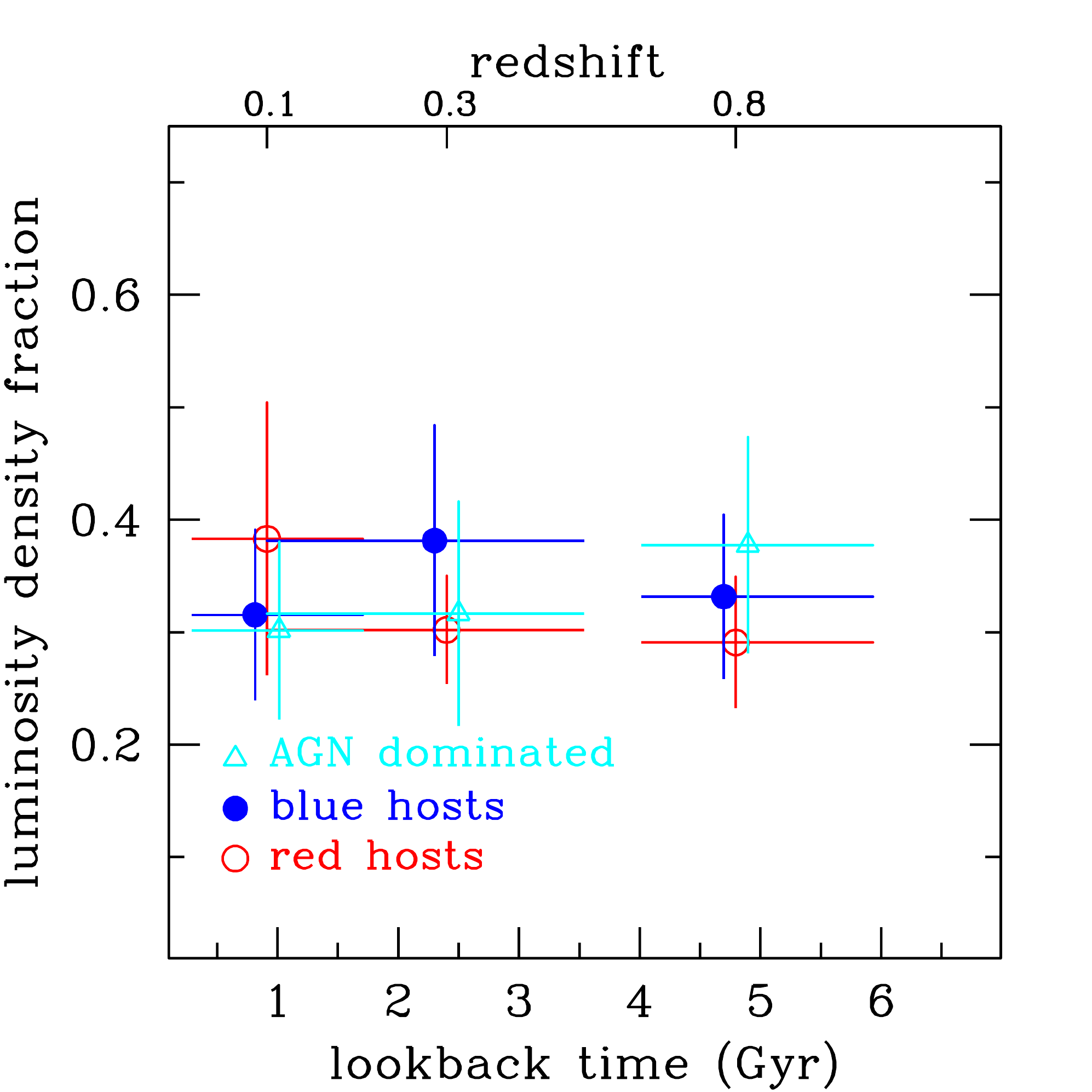}
\end{center}
\caption{The fraction of the  X-ray luminosity density associated with
  different AGN samples  is plotted as a function  of lookback time
  (lower x-axis) and redshift (upper x-axis).  The red
  open  circles  are  for  red  AGN hosts,  the  blue  filled  circles
  represent AGN in blue hosts and the cyan triangles correspond to AGN
  with optical  colours dominated by the central  engine. The vertical
  errorbars are Poisson estimates propagated from the uncertainties in
  the X-ray  luminosity density.  The horizontal  errors represent the
  redshift interval probed by  the different samples.  For clarity the
  filled  circles and  triangles  are offset  by  -0.2 and  +0.2\,Gyrs
  respectively.}\label{fig_ld_fraction}
\end{figure}

More relevant  in the study  of the dominant  AGN accretion mode  as a
function of  redshift is  the fraction of  the accretion power  of the
Universe that  is associated with red/blue hosts  at different epochs.
Figure \ref{fig_ld} plots the hard  band X-ray luminosity density as a
function of  luminosity for the three  samples used in  this paper. At
each redshift bin  we sum up separately the  luminosity density of AGN
in  red/blue hosts  and X-ray  sources with  colours dominated  by the
central  engine and  divide with  the total  luminosity  density.  The
results are  plotted as  a function of  lookback time and  redshift in
Figure  \ref{fig_ld_fraction}.   Within  the  uncertainties  there  is
little change with redshift of  the relative fraction of the accretion
power in red/blue  AGN hosts from $z=0.8$ to  $z=0.1$.  This statement
is quantified  by fitting  a straight line  to the  luminosity density
fraction versus  redshift datapoints of  Figure \ref{fig_ld_fraction}.
For the  slope of the  line we  find best fit  values and 68  per cent
confidence level  errors ($\Delta\chi^2=1$) of $+0.01^{+0.14}_{-0.15}$
(blue AGN  hosts) and $-0.05^{+0.13}_{-0.15}$ (red  AGN hosts).  Based
on the  uncertainty in the  slope of the  linear fit we  then estimate
that  at the  68  per  cent confidence  level  the luminosity  density
associated with blue AGN hosts cannot increase or decrease more than a
factor of about 1.4 between $z=0.8$  and $z=0.1$. For AGN in red hosts
we find that their luminosity density cannot increase or decrease more
than factors of 1.6 and  1.2 respectively, from $z=0.8$ to $z=0.1$ (68
per  cent confidence  level).  We  also  caution that  the points  for
red/blue AGN hosts in Figure \ref{fig_ld_fraction} are lower limits as
the typical host galaxy colours  of broad-line QSOs are still not well
constrained.

\subsection{The optical luminosity functions of AGN and galaxies}
\begin{figure}
\begin{center}
\includegraphics[height=0.9\columnwidth]{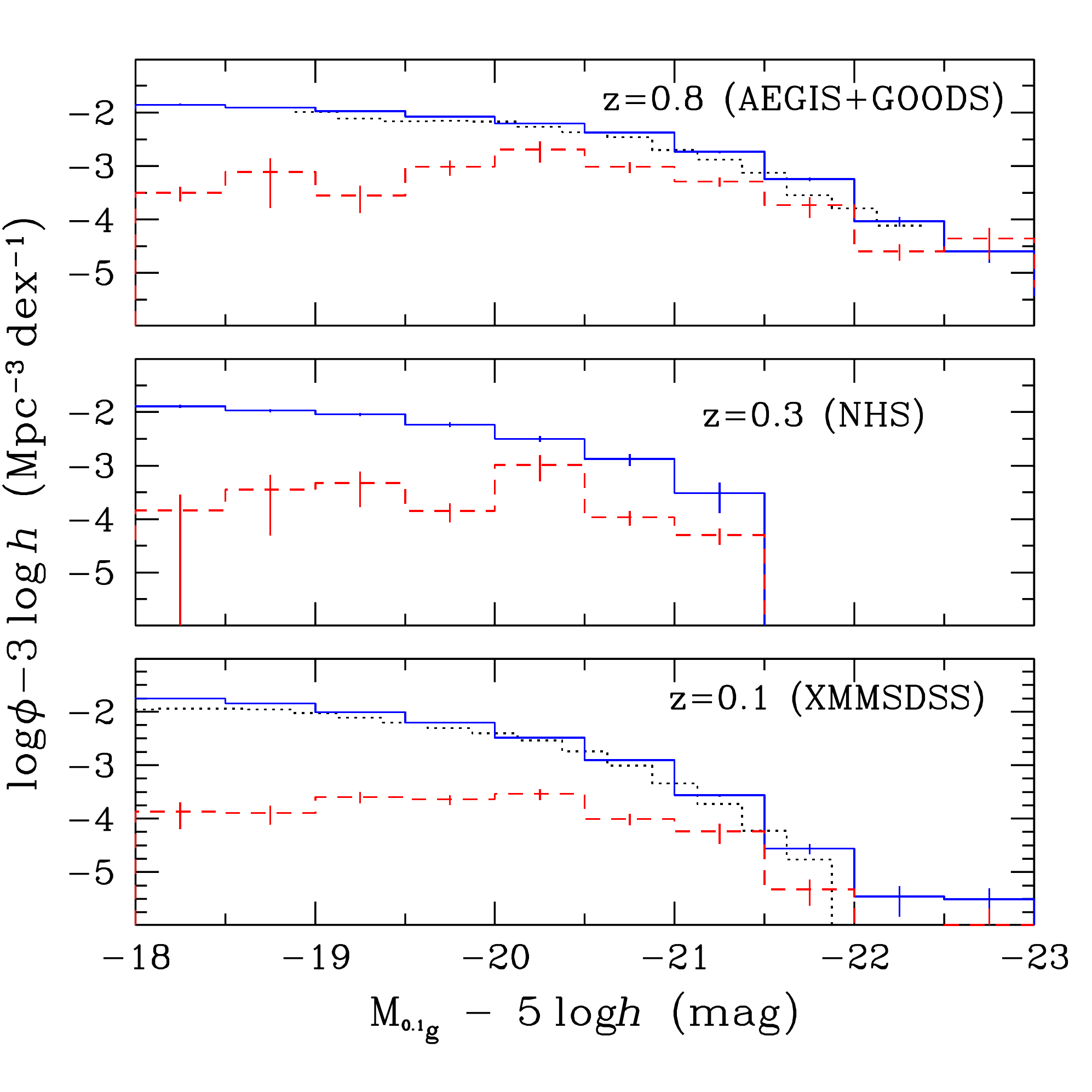}
\end{center}
\caption{Optical luminosity function for galaxies and X-ray AGN in the
  XMM/SDSS,  NHS and  AEGIS/GOODS samples  in  $M_{^{0.1}g}$ magnitude
  bins. The red  dashed histogram corresponds to X-ray  AGN.  The blue
  solid lines are for the  overall galaxy population.  These should be
  compared  with the  dotted black  histograms which  are the  OLFs of
  galaxies estimated  by Blanton et al. (2007)  at redshifts intervals
  similar to those  of the XMM/SDSS and the  AEGIS galaxy samples. For
  the NHS at  $z\approx0.3$ there is no estimate of  the galaxy OLF in
  $M_{^{0.1}g}$  bins   with  which   we  can  compare   directly  our
  results.}\label{fig_olf}
\end{figure}

\begin{figure}
\begin{center}
\includegraphics[height=0.8\columnwidth]{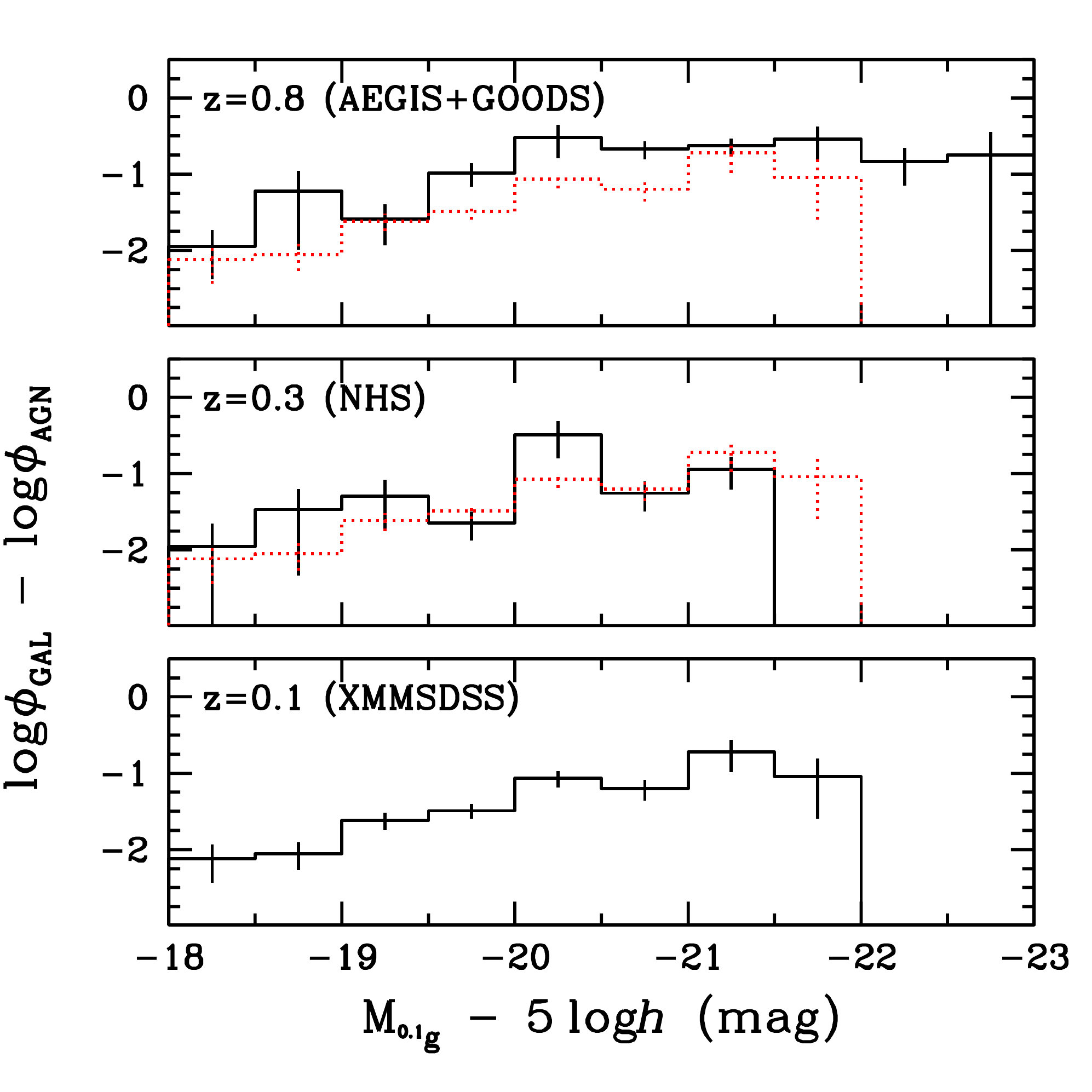}
\includegraphics[height=0.8\columnwidth]{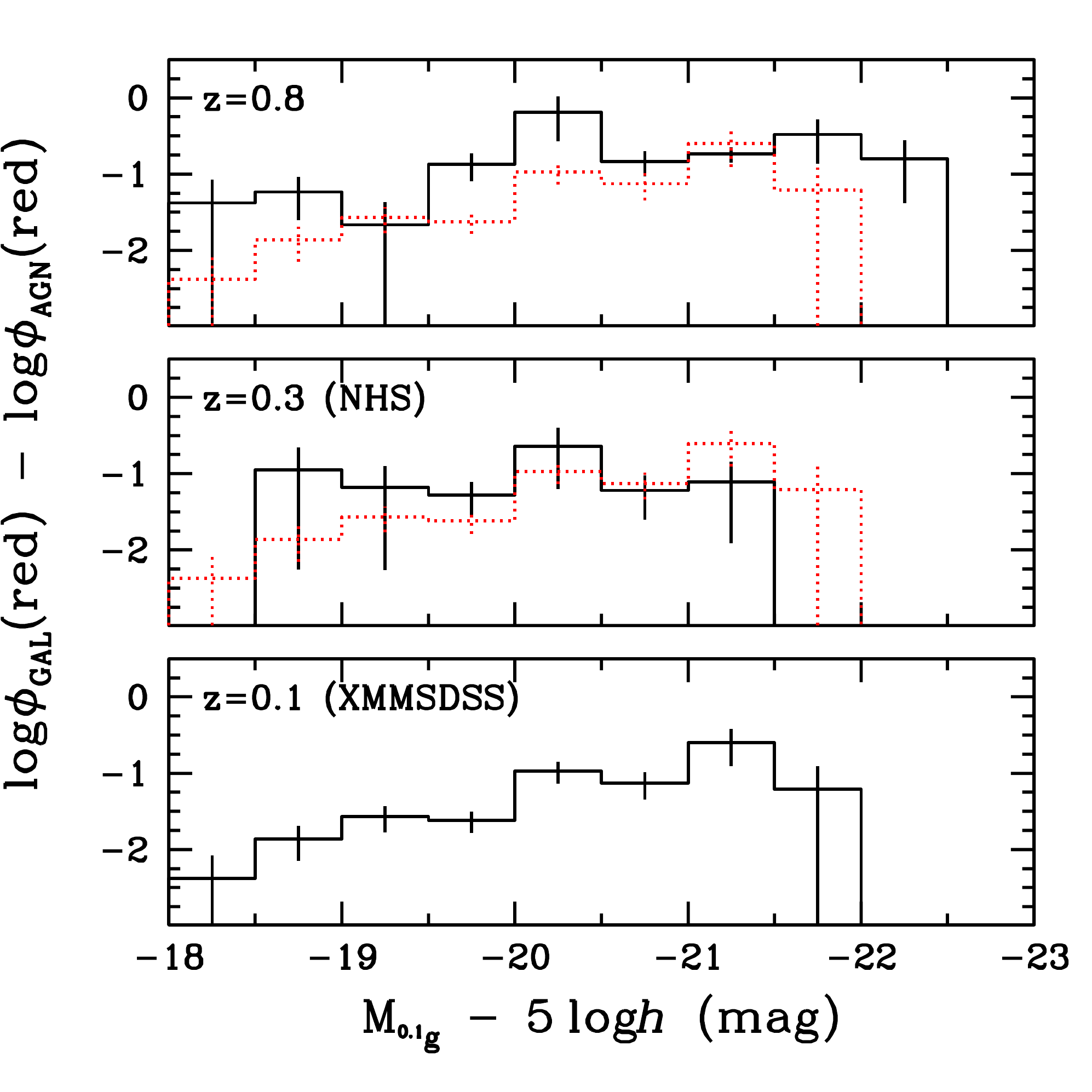}
\includegraphics[height=0.8\columnwidth]{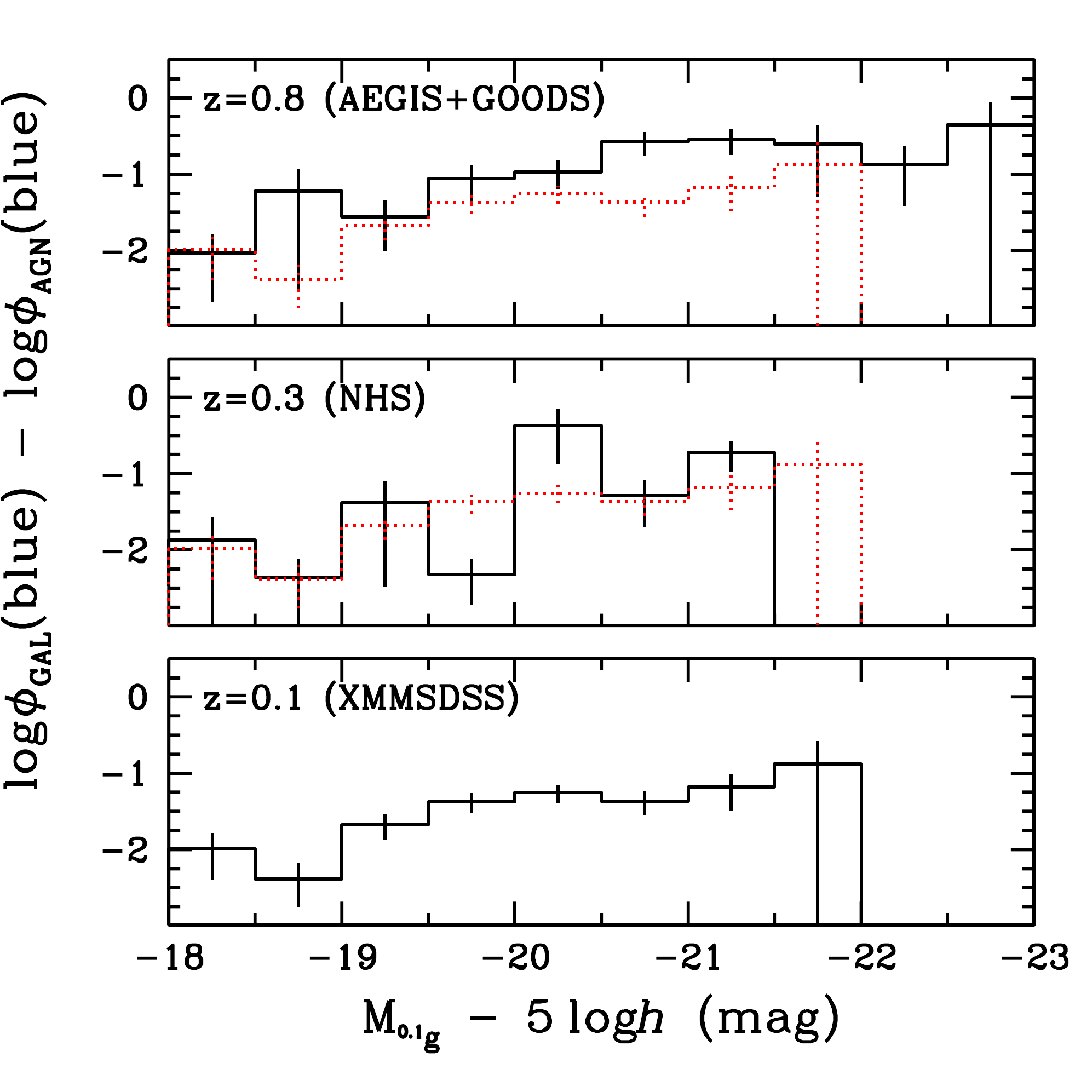}
\end{center}
\caption{Ratio of the optical  luminosity function X-ray AGN over that
  of galaxies plotted  as a function $M_{^{0.1}g}$. In  each panel the
  XMM/SDSS, NHS  and AEGIS/GOODS samples are  plotted separately.  The
  top panel is for the total sample, the middle is the fraction of red
  cloud AGN relative to red  galaxies. The bottom panel plots blue AGN
  relative to  blue galaxies.  For comparison, in  each panel  the red
  dashed histogram shown in  the $z=0.3$ and $z=0.8$ plots corresponds
  to  the $z\approx0.1$  X-ray  fraction estimated  from the  XMM/SDSS
  survey.  The errorbars  are  Poisson estimates  propagated from  the
  uncertainties in  the optical luminosity functions of  X-ray AGN and
  galaxies.  }\label{fig_olf_fraction}
\end{figure}

We also  explore changes  with redshift of  the fraction of  X-ray AGN
relative to  the overall galaxy  population. For that we  estimate the
optical luminosity function of galaxies and X-ray AGN in $M_{^{0.1}g}$
absolute magnitude bins.  X-ray AGN with optical continua dominated by
the  central engine  are excluded  from the  OLF calculation  as their
$M_{^{0.1}g}$ may not be  dominated by stellar light.  Including those
sources  in   the  analysis  however,   would  not  change   the  main
conclusions.  The  results are  presented in Figure  \ref{fig_olf} and
Table \ref{tab_OLF}.  For comparison,  also plotted in that figure are
the galaxy  OLFs estimated in $M_{^{0.1}g}$ bins  at $z\approx0.1$ and
$z\approx0.8$ by \cite{Blanton2006} using  SDSS-DR4 data and the first
data  release of  the  DEEP2 spectroscopic  survey respectively.   Our
galaxy  OLFs   are  in  good   agreement  with  that   study.   Figure
\ref{fig_olf_fraction} plots the ratio of  the X-ray AGN OLF over that
of the overall  galaxy population.  Within the errors  the fraction of
X-ray  AGN remains  roughly  constant with  redshift  in the  absolute
magnitude range  shown in the figure.   This result holds  if we split
the galaxy and AGN samples  by rest-frame colour. The $\chi^2$ test is
employed to explore in  a quantitative way differences between $z=0.1$
and  $z=0.8$ in  the  $M_{^{0.1}g}$ distribution  of  the fraction  of
AGN. The $\chi^2$ statistic is

\begin{equation} 
\chi^2 = \sum_{i} \left ( \frac{f_{i,z=0.1} -
  f_{i,z=0.8}}{\sigma_{i}} \right )^2,
\end{equation}

\noindent where $f_{i,z=0.1}$ and $f_{i,z=0.8}$ is the fraction of AGN
in the  optical luminosity  bin $i$ at  redshifts $z=0.1$  and $z=0.8$
respectively, and  $\sigma_{i}$ is  the combined error  which accounts
for the  uncertainty in both $f_{i,z=0.1}$ and  $f_{i,z=0.8}$. For the
total X-ray sample  we estimate $\chi^2 =15.6$ (8  degrees of freedom)
which corresponds to a probability $P=0.95$ that the two distributions
are different.  For the red  and blue AGN subsamples the corresponding
probabilities  are $P=0.83$  ($\chi^2 =11.7$,  8 d.o.f.)  and $P=0.89$
respectively  ($\chi^2  =13.1$,  8  d.o.f.). We  therefore  find  only
marginal evidence, significant  at best at the 95  per cent confidence
level  ($2\sigma$  for  a  normal  distribution), for  a  change  with
redshift  of the  X-ray AGN  fraction  in a  given optical  luminosity
bin.  This is surprising  given that  the space  density of  X-ray AGN
drops by almost 1\,dex since $z\approx0.8$.

\subsection{The stellar mass functions of AGN and galaxies}
\begin{figure}
\begin{center}
\includegraphics[height=0.9\columnwidth]{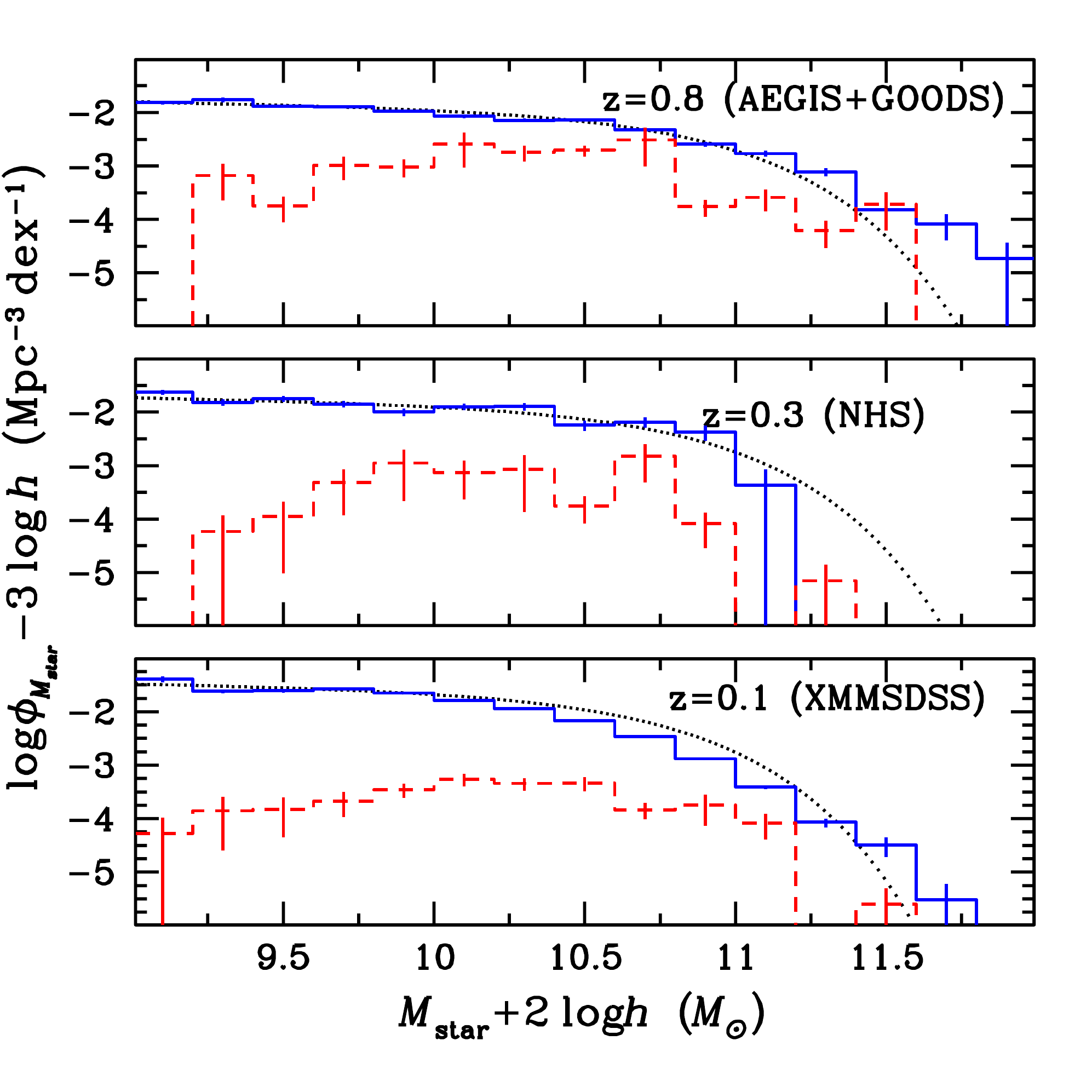}
\end{center}
\caption{Stellar  mass  function of  galaxies  and  X-ray  AGN in  the
  XMM/SDSS,  NHS and  AEGIS/GOODS  samples. The  red dashed  histogram
  corresponds to X-ray  AGN. The blue solid lines  are for the overall
  galaxy population.   These should be compared with  the dotted black
  curves  which  are the  Schechter  fits  to  the mass  functions  of
  galaxies at  $z=0$, $z=0.3$  and $z=0.7$ presented  by Borch  et al.
  (2006) after correcting for the different Hubble constant adopted in
  their paper.}\label{fig_mf}
\end{figure}

\begin{figure}
\begin{center}
\includegraphics[height=0.8\columnwidth]{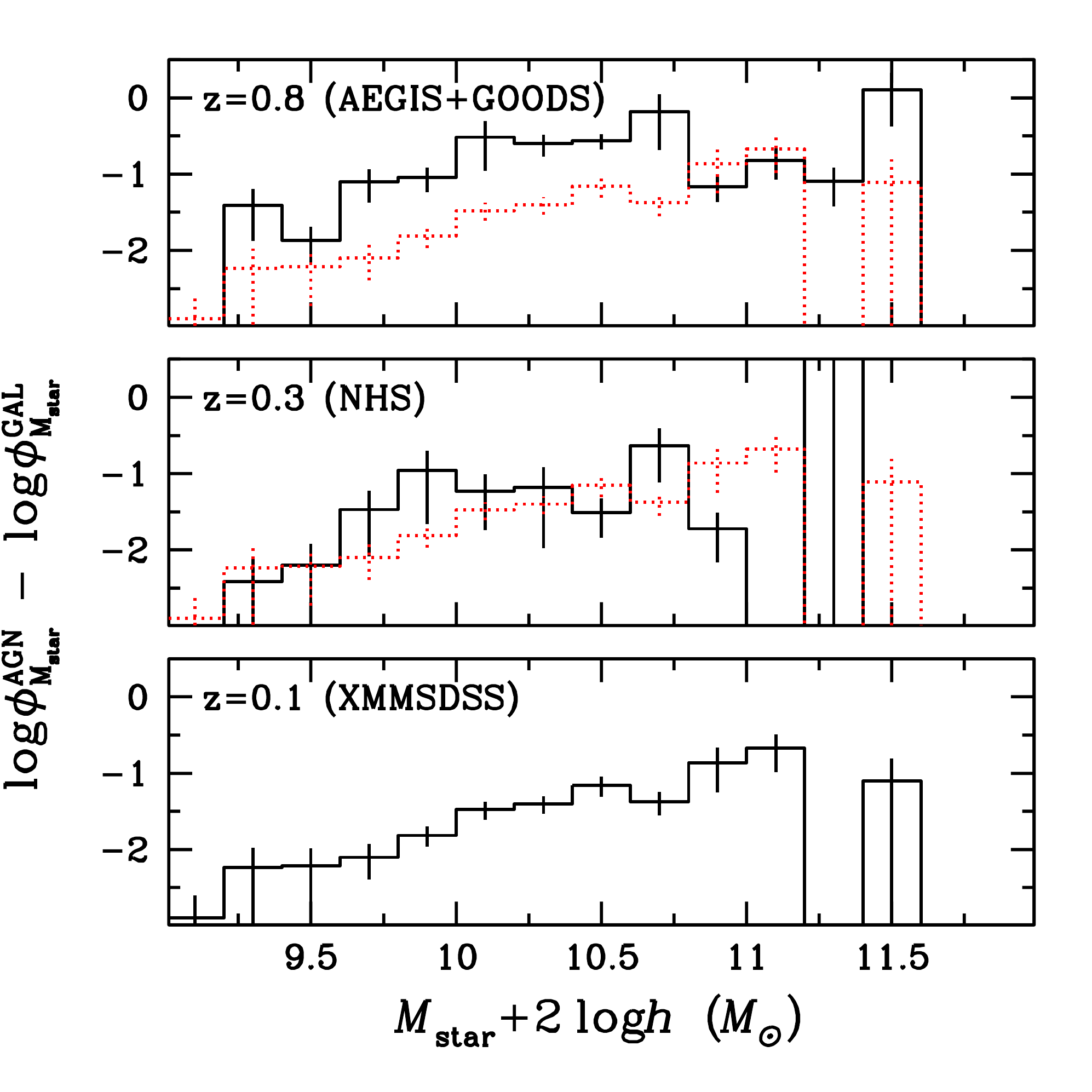}
\includegraphics[height=0.8\columnwidth]{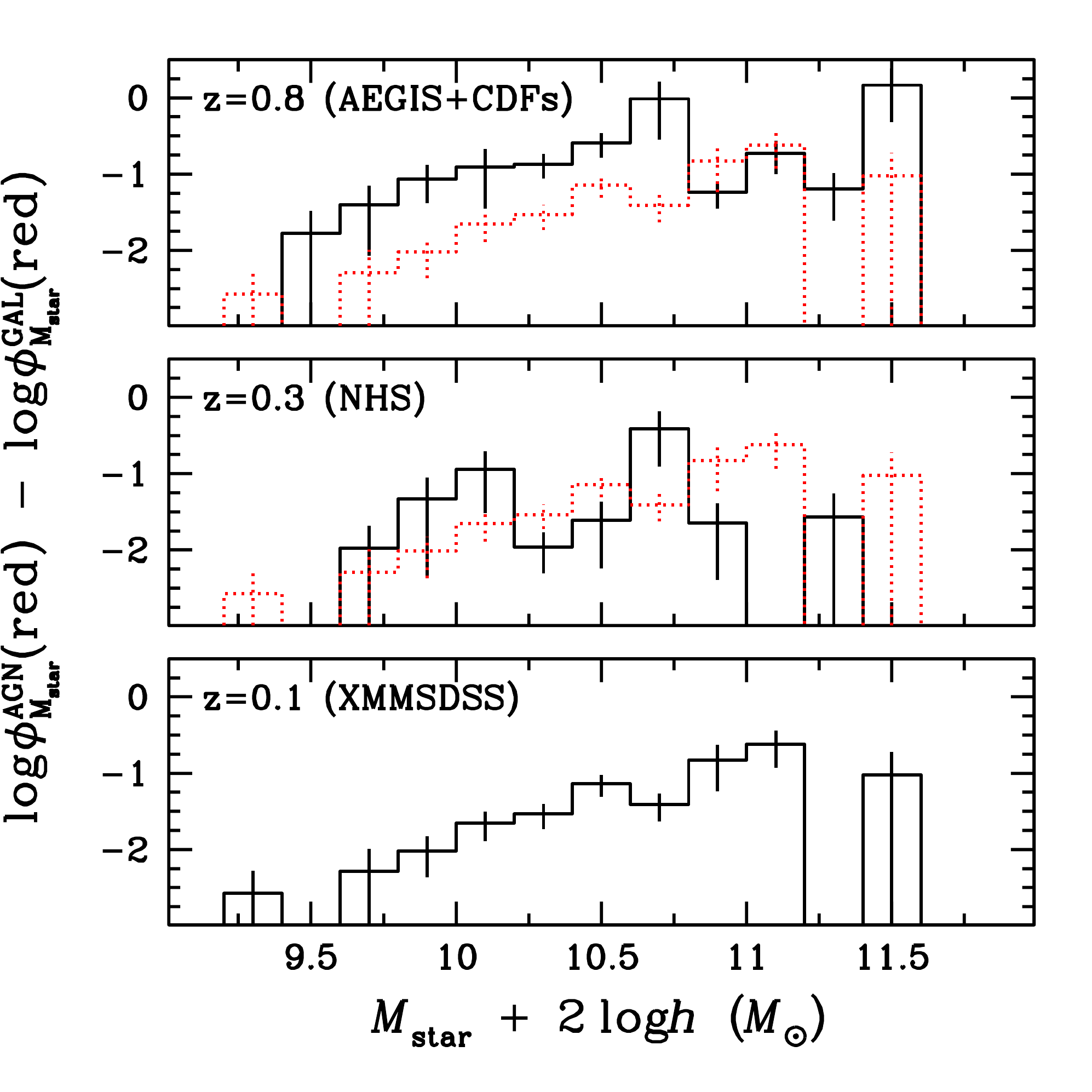}
\includegraphics[height=0.8\columnwidth]{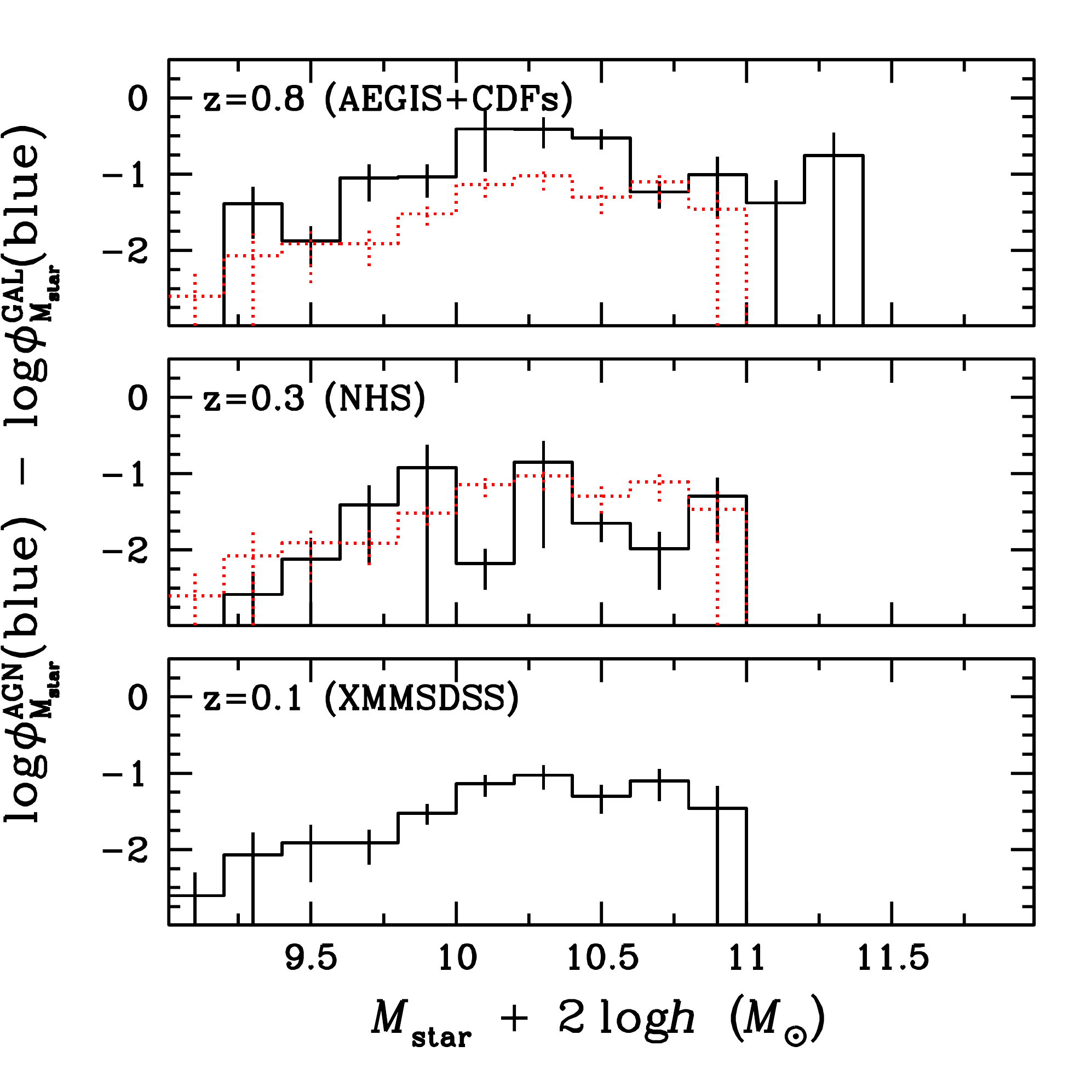}
\end{center}
\caption{Ratio of the mass function of X-ray AGN over that of galaxies
  plotted against  stellar mass. In  each panel the XMM/SDSS,  NHS and
  AEGIS/GOODS samples  are plotted separately.   The top panel  is for
  the total sample. The middle panel  is the fraction of red cloud AGN
  relative to red galaxies.  The  bottom panel plots blue AGN relative
  to  blue galaxies.  For comparison,  in  each panel  the red  dashed
  histogram shown in the $z=0.3$  and $z=0.8$ plots corresponds to the
  $z\approx0.1$ X-ray fraction estimated from the XMM/SDSS survey. The
  errorbars are Poisson estimates propagated from the uncertainties in
  the    mass     functions    of    X-ray     AGN    and    galaxies.
}\label{fig_mf_fraction}
\end{figure}

The conclusions  are different if  the fraction of galaxies  that host
X-ray  AGN is plotted  against stellar  mass.  In  this case  there is
evidence for evolution in the  sense that at higher redshifts a larger
fraction  of galaxies  host X-ray  AGN at  a given  stellar  mass bin.
Figure \ref{fig_mf} and Table  \ref{tab_MF} present the mass functions
of galaxies  and X-ray  AGN at redshifts  $z\approx0.1$, 0.3  and 0.8.
X-ray  AGN  with optical  continua  dominated  by  the central  engine
(i.e. broad emission lines and/or  a prominent nuclear point source in
their  optical image)  are excluded  from the  MF  estimation.  Figure
\ref{fig_mf} also shows the  galaxy MFs of \cite{Borch2006} at similar
redshifts.  There  is fair agreement with our  MF estimates suggesting
that our  simple approach for  calculating stellar masses  is adequate
for  statistical   studies.   The   agreement  with  the   results  of
\cite{Borch2006} is  also good if  the MF is  split into red  and blue
galaxies.  Figure  \ref{fig_mf_fraction} plots the ratio  of X-ray AGN
MF  over that  of  galaxies.  In  all  three samples  the fraction  of
galaxies that host AGN increases  with stellar mass. The steepest rise
is  for the  low  redshift sample  whereas  in both  the  NHS and  the
AEGIS+CDFs samples the fraction of  X-ray AGN appears to flatten above
about $\rm 10^{10} \, M_{\odot}$.  Also, the fraction of galaxies that
host  X-ray  AGN  drops  with  decreasing  redshift  from  $z=0.8$  to
$z=0.1$. Using  the $\chi^2$ statistical  test to compare  the $z=0.8$
with $z=0.1$ AGN fraction distribution we find $\chi^2=31.5$ (9 d.o.f)
and a  probability $P=99.98$  per cent ($3.7\sigma$  in the case  of a
normal distribution)  that the two distributions  are different.  This
is the result of the building up of the mass function of galaxies with
time and the overall decline of  the space density of X-ray AGN to low
redshift.  At  stellar masses $\ga 10^{11} \,  _{\odot}$, although the
statistics are poor, there is  tentative evidence that the fraction of
X-ray  AGN is  not  a strong  function  of redshift.   This is  likely
related  to the  fact that  the mass  function above  this  mass limit
evolves      little     with     redshift      since     $z\approx0.8$
\citep{Borch2006,Perez2008}.     Figure   \ref{fig_mf_fraction}   also
presents the  fraction of  X-ray AGN in  red/blue hosts.   The overall
trends are similar  to the full galaxy and  AGN populations. Using the
$\chi^2$  test the  probabilities that  the  X-ray AGN  fraction as  a
function of stellar mass changes  from $z=0.8$ to $z=0.1$ are $P=0.98$
($\chi^2 =19.3$,  9 d.o.f.)   for the red  AGN subsample  and $P=0.97$
($\chi^2 =18.8$, 9 d.o.f.) for the blue AGN subset.

\section{Discussion}

A slew of  AGN fueling modes are proposed  in the literature including
hot gas accretion (Croton et al. 2006), major galaxy mergers (e.g.  Di
Matteo  et al.   2005), internal  instabilities or  minor interactions
(e.g.  Hopkins \&  Hernquist 2006) and accretion of  recycled gas from
evolved stars  (e.g.  Ciotti  \& Ostriker 1997,  2007).  Semi-analytic
models  use combinations of  these physical  processes to  explain the
formation  of  galaxies and  the  growth  of  SMBHs at  their  centres
\citep[e.g. ][]{Somerville2008, Fontanot2007, Fanidakis2011}.  In some
of  these models  the evolution  of the  AGN population  is intimately
related  to   changes  in  the  dominant  accretion   mode  with  time
\citep[e.g. ][]{Fanidakis2011}.  Such  changes are expected to imprint
on the properties of AGN hosts, e.g.  their morphology, star-formation
history, stellar mass and small/large scale environment.

In this  paper the rest-frame colours  of AGN hosts are  used as proxy
for their  integrated star-formation history to  explore variations of
this  quantity with  redshift  from $z=0.8$  to  $z=0.1$.  Within  the
uncertainties, we  do not  find evidence that  the relative  number of
red/blue  X-ray AGN  hosts changes  with redshift,  in  agreement with
recent results from ChaMP  \citep{Haggard2010}.  Also, the fraction of
the accretion density associated with active SMBHs in the red sequence
and the blue  cloud is nearly constant since  $z=0.8$.  These findings
suggest that the galaxies that  host X-ray AGN have similar integrated
star-formation  histories  at  all  redshifts probed  in  this  paper,
$z=0.1$, 0.3  and 0.8. This  argues against systematic changes  of the
dominant accretion mode from $z=0.8$  to $z=0.1$.  The same process or
combination  of processes for  fueling SMBHs  are likely  in operation
from $z=0.8$ to the present day.

There  are  caveats  in  this  conclusion. A  large  fraction  of  the
accretion power at  all redshifts is associated with  luminous AGN. In
many of those sources the  radiation from the central engine dominates
the optical  continuum. The  optical colours of  those systems  do not
provide  any  information on  the  stellar  population  of their  host
galaxies.   Our  analysis   is  therefore  insensitive  to  systematic
variations  with redshift  in  the  fueling mode  of  broad line  AGN.
Additionally our conclusions are for moderate luminosity AGN, $L_X \rm
(2-10\,keV)=10^{41}-10^{44} \, erg \, s^{-1}$ and do not apply to QSOs
with  luminosities $\rm  L_X>10^{44} \rm  \,  erg \,  s^{-1}$, as  the
volumes of all three samples used here are too small to identify large
numbers of those sources. Scattered AGN light may also contaminate the
observed colours  of galaxies,  even in the  absence of  broad optical
emission lines or a dominant nuclear point sources.  Nevertheless, the
work  of  \cite{Pierce2010}  suggests   that  this  effect  is  small.
Reddening    of    the    intrinsic    galaxy    colours    by    dust
\citep{Cardamone2010} is also expected to affect our conclusions.

If  there  is  little or  no  change  with  redshift in  the  relative
importance of SMBH  fueling modes, how is the  strong evolution of the
AGN  population  explained?   An   important  piece  of  evidence  for
addressing this question is the  nearly constant fraction of X-ray AGN
relative  to the  OLF of  galaxies as  a function  of  redshift.  This
suggest that the decline of  the accretion power of the Universe since
$z\approx1$ is  linked, at  least to the  first approximation,  to the
evolution of  the galaxy OLF, which  in turn is the  result of secular
evolution processes  and the decrease with time  of the star-formation
rate density  of the Universe.   Our results are  therefore consistent
with  a  fixed accretion  mode  (or  combination  of accretion  modes)
superimposed on an evolving galaxy population.

The  stellar  mass  function  of  AGN relative  to  galaxies  provides
complementary  information on the  evolution of  active SMBHs.   It is
found that X-ray  AGN hosts out to $z\approx0.8$ span  a wide range of
stellar masses, although their fraction increases with stellar mass in
agreement   with  previous   studies   \citep{Kauffmann2003,  Shi2008,
  Silverman2009_SFR,  Xue2010,  Haggard2010}.    This  trend  is  more
pronounced at  lower redshift, where the distribution  of the fraction
of galaxies with X-ray AGN increases nearly monotonically with stellar
mass.   At higher redshifts  however, the  distribution levels  off at
intermediate masses, $\rm \ga 10^{10}\,M_{\odot}$.  Similar trends are
reported by \cite{Shi2008}  and \cite{Bundy2008}, although their X-ray
AGN samples  are selected at brighter luminosities,  $L_X> \rm 10^{42}
\, erg \, s^{-1}$, compared to our limit of $L_X \rm (2-10\,keV) > \rm
10^{41} \,  erg \,  s^{-1}$.  It  is also found  that the  fraction of
galaxies hosting  an active  SMBH decreases with  redshift at  a given
stellar mass bin.  For the red AGN/galaxy populations this decrease is
because of the  build-up of the red galaxy  stellar mass function from
$z=0.8$ to $z=0.1$  and the overall decline of  the AGN space density.
For AGN with  blue colours however, the blue  galaxy MF hardly evolves
with redshift \citep[e.g.][]{Borch2006}.   The decline of the fraction
of  blue  galaxies that  host  AGN  at a  given  stellar  mass bin  is
primarily because the space density of blue AGN drops with redshift.

\cite{Kauffmann_Heckman2009}  found evidence for  two regimes  of SMBH
growth  in the nearby  Universe. The  first (``feast'')  is associated
with  actively  star-forming  galaxies   and  is  characterized  by  a
log-normal distribution of accretion rates  peaked at a few percent of
the Eddington limit.  For these galaxies the gas supply is abundant to
feed  both star-formation  and  substantial SMBH  growth.  The  second
regime  (``famine'') is  associated with  evolved galaxies  and  has a
power-law distribution of accretion rates.  The SMBHs of these systems
is  likely  to  be  fed  by  recycled gas  from  the  evolved  stellar
population and  their growth rates  are small. The  transition between
the  two modes  occurs where  the inverse  of  specific star-formation
rate, $\rm  M_{star}/SFR$, is comparable  to the age of  the Universe,
i.e. for galaxies that  experience significant growth of their stellar
mass today.  Independent  work has shown that at  a given stellar mass
interval  the quantity  $\rm M_{star}/SFR$  decreases  with increasing
redshift \citep[e.g.][]{Perez2008}, i.e.  galaxies were more active in
the  past.  Therefore  at a  given stellar  mass bin  the  fraction of
galaxies above the $\rm M_{star}/SFR$ limit which marks the transition
from the  ``famine'' to the  ``feast'' SMBH growth regimes  drops with
redshift.   If the X-ray  AGN selection  is dominated  by SMBH  in the
``feast'' mode then  the observed decline of the  fraction of galaxies
that host X-ray  AGN from $z=0.8$ to $z=0.1$ at  a given stellar mass,
could be  a direct result of  the evolution of  the $\rm M_{star}/SFR$
with redshift, i.e. the decline of the volume density of potential AGN
hosts.

It is indeed  expected that most sources in  our X-ray selected sample
will  be in  the ``feast''  mode.  AGN  in the  ``famine''  regime are
likely to be  very faint, close to or below  the luminosity limit $L_X
\rm          (2-10\,keV)=10^{41}           \,          erg          \,
s^{-1}$. \cite{Kauffmann_Heckman2009}  showed that SMBHs  that grow in
the  ``famine'' regime  live in  old galaxies  with  4000\,\AA\, break
amplitude $D_n  \rm (4000)>1.8$.  AGN in such  galaxies typically have
[OIII]\,5007   emission  line   luminosities,   $L_{[\rm  OIII]}   \la
10^{6}-10^{7}\,L_\odot$  (see Figure  12 of  Kauffmann et  al.  2003).
Adopting  a  ratio between  X-ray  and [OIII]\,5007\AA\,  luminosities
$\log  (L_X  (  {  \rm  2-10  \,  keV  }  )  /  L_{\rm  [OIII]})=1.66$
(Georgantopoulos \& Akylas 2009), we estimate $L_X \rm ( 2-10 \, keV )
\la 10^{41}-10^{42} \, erg \,  s^{-1}$ for the optically selected SDSS
AGN in the ``famine'' regime  ($D_n \rm (4000)>1.8$).  We caution that
there is a large scatter in the $L_X/ L_{\rm [OIII]}$ ratios and faint
or obscured  AGN may have much  lower ratios than what  is adopted here
\citep[e.g.][]{Netzer2006,      Georgantopoulos_Akylas2010}.       The
conversion  of  [OIII]\,5007\AA\,  to  X-ray luminosity  is  therefore
uncertain and the X-ray luminosities  above may be overestimated by as
much as 1\,dex.   In any case these calculations  show that most X-ray
AGN with $L_X \rm  ( 2-10 \, keV ) > 10^{41} \,  erg \, s^{-1}$ are in
the ``feast''  stage of  SMBH growth.  This  is in agreement  with the
results  of \cite{Silverman2009_SFR}  who  measured 4000\,\AA\,  break
amplitudes $D_n \rm (4000)<1.8$ for most X-ray AGN at $z\approx1$ with
$L_X \rm ( 0.5-10  \, keV ) > 10^{42} \, erg  \, s^{-1}$ in the COSMOS
survey.

A  number  of  studies  already  point  to  a  link  between  AGN  and
star-formation  out  to  $z\approx1$.   The similarities  between  the
global        star-formation       rate        density       evolution
\citep[e.g.][]{Hopkins_A2004} and  the accretion power  history of the
Universe \citep[e.g.][]{Aird2010} suggests  that the two processes are
connected, at least in a statistical manner \citep[e.g.][]{Zheng2009}.
There is also  evidence that the mean star-formation  rate of moderate
luminosity X-ray AGN hosts evolves  with time in a manner that closely
mirrors   the  overall  galaxy   population  \citep{Silverman2009_SFR,
  Lutz2010, Shao2010}.  This similarity  would imprint in the $g$-band
OLF  of  AGN  and galaxies,  as  found  in  this paper.   In  apparent
contradiction to those results, \cite{Bundy2008} found evidence for an
association   between  AGN   activity   and  the   quenching  of   the
star-formation  in  galaxies,  i.e.   their transformation  from  blue
star-forming  to   evolved  red  systems.   However,   this  does  not
necessarily oppose studies that claim  a link between the formation of
young  stars  and  the growth  of  SMBHs,  as  long as  the  quenching
timescale  is  comparable  to  or  longer than  the  duration  of  the
star-formation burst.

Our analysis  suggests that the  evolution of the  moderate luminosity
X-ray  AGN population is  related to  (i) the  presence of  a specific
star-formation  rate  of  galaxies  which marks  the  transition  from
abundant to  little gas supply to  the SMBH and (ii)  the decline with
redshift of the mean specific star-formation rate of galaxies. In this
interpretation  it is  the decrease  at lower  redshift of  the number
density of potential  AGN hosts that drives the  evolution of moderate
luminosity  X-ray  AGN since  $z=0.8$.   This  further underlines  the
importance  of constraining galaxy  formation scenarios  to understand
the growth of SMBHs.

\section{Conclusions}

Deep small  angular size  surveys are combined  with wide-area/shallow
samples to study  the evolution of moderate luminosity  X-ray AGN from
$z=0.1$ to  $z=0.8$. The rest-frame colours  of AGN hosts  are used as
proxy for their integrated star-formation history.

It is found that the fraction of the accretion density associated with
red  or  blue AGN  hosts  does not  change  with  redshift within  the
uncertainties.  This  argues against  scenarios in which  the dominant
mode of SMBH growth changes with redshift since $z=0.8$.

There  is  evidence  that  the  evolution of  the  AGN  population  is
associated with  the decrease  with redshift of  the space  density of
potential AGN hosts. We find that  the fraction of galaxies of a given
$g$-band absolute magnitude that host X-ray AGN changes little (95 per
cent  confidence level)  from $z=0.8$  to $z=0.1$,  despite  the rapid
decline of the  AGN space density at this  redshift interval (close to
1\,dex).  This suggests  a link between AGN, the  secular evolution of
galaxies  and the  decline of  the global  star-formation rate  to the
present day.

When the  fraction of  AGN among galaxies  is plotted  against stellar
mass  a systematic  decline with  redshift  is found  (99.98 per  cent
confidence level).   This is  because of the  build-up of  the stellar
mass of galaxies and the decrease of the space density of AGN to lower
redshift.

It is  argued that  these trends are  consistent with a  picture where
most X-ray AGN live in  systems with abundant gas reservoirs, but
as the gas supply, and hence the specific star formation of the galaxy
population,  drops  with redshift,  the  number  density  of AGN  also
decreases.

\begin{table*}
\caption{X-ray Luminosity Function and X-ray Luminosity Density}\label{tab_XLF}
\begin{center} 
\scriptsize
\begin{tabular}{c c c  c c c c}

\hline
$\log L_X(\rm 2-10\,keV)$ &  
$\phi(ALL)$ &
$\phi(red)$ &
$\phi(blue)$ &

$L_X\,\phi(ALL)$ &
$L_X\,\phi(red)$ &
$L_X\,\phi(blue)$ 
\\

($\rm erg\,s^{-1}$) &
($\rm 10^{-4}\,Mpc^{-3}$) &
($\rm 10^{-4}\,Mpc^{-3}$) &
($\rm 10^{-4}\,Mpc^{-3}$) &

($\rm 10^{38}\,erg\,s^{-1}\,Mpc^{-3}$) &
($\rm 10^{38}\,erg\,s^{-1}\,Mpc^{-3}$) &
($\rm 10^{38}\,erg\,s^{-1}\,Mpc^{-3}$) 
\\
\hline
\multicolumn{7}{c}{XMM/SDSS}\\ \hline

41-42 & $4.3\pm0.7$  & $2.7\pm0.5$     &   $1.5\pm0.4$       &  $1.1\pm0.2$  & $0.6\pm0.1$ & $0.5\pm0.1$\\
42-43 & $1.5\pm0.2$  & $0.4\pm0.1$     &   $0.8\pm0.2$       &   $4.9\pm0.7$ & $1.2\pm0.3$ & $2.1\pm0.5$\\
43-44 & $0.14\pm0.06$ & $0.08\pm0.06$  &   $0.012\pm0.009$   &    $2.7\pm1.0$  & $1.5\pm0.9$ & $0.2\pm0.1$\\
\hline

\multicolumn{7}{c}{NHS}\\ \hline
41-42 & $9.6\pm3.5$  & $4.6\pm2.3$       &  $5.0\pm2.6$          & $3.3\pm1.1$ & $1.9\pm0.3$  & $1.5\pm0.7$ \\
42-43 & $1.9\pm0.4$  & $0.8\pm0.2$       &   $0.9\pm0.3$         & $6.2\pm1.0$ & $2.3\pm0.1$  & $2.7\pm0.7$ \\
43-44 & $0.24\pm0.06$ & $0.04\pm0.02$   &    $0.07\pm0.3$   & $6.2\pm1.7$ & $0.6\pm0.2$  & $1.8\pm0.9$ \\
\hline

\multicolumn{7}{c}{AEGIS/GOODS}\\ \hline

41-42 & $21.2\pm5.7$   & $9.4\pm4.3$       &  $11.7\pm3.7$       & $5.3\pm1.1$ & $2.5\pm0.8$  & $2.8\pm0.8$ \\
42-43 & $5.4\pm0.7$     & $2.9\pm0.4$       &   $2.4\pm0.5$        & $20.0\pm3.4$ & $10.6\pm2.5$  & $9.0\pm2.3$ \\
43-44 & $1.1\pm0.2$     & $0.31\pm0.07$   &    $0.4\pm0.1$       & $31.9\pm4.6$ & $7.1\pm1.9$  & $9.9\pm2.6$ \\
44-45 & $0.12\pm0.05$ &  $0.02\pm0.02$  & $0.01\pm0.01$      & $17.7\pm6.5$ & $1.6\pm1.6$ & $3.0\pm3.0$\\
\hline

\end{tabular} 
\begin{list}{}{}
\item 
The columns  are: (1): X-ray  luminosity interval; (2):  space density
for all X-ray  AGN; (3) space density for X-ray AGN  in red hosts; (4)
space  density for  X-ray  AGN  in blue  hosts;  (5) X-ray  luminosity
density for all X-ray AGN;  (6) X-ray luminosity density for X-ray AGN
in  red hosts;  (7) X-ray  luminosity density  for X-ray  AGN  in blue
hosts;
\end{list}
\end{center}
\end{table*}

\begin{table*}
\caption{Optical Luminosity Function of AGN and galaxies}\label{tab_OLF}
\begin{center} 
\scriptsize
\begin{tabular}{c c c  c c c c}

\hline
$M_{^{0.1}g}$ &  
$\phi_{AGN}(ALL)$ &
$\phi_{AGN}(red)$ &
$\phi_{AGN}(blue)$ &

$\phi_{gal}(ALL)$ &
$\phi_{gal}(red)$ &
$\phi_{gal}(blue)$ 
\\

(mag) &
($\rm 10^{-4}\,Mpc^{-3}$) &
($\rm 10^{-4}\,Mpc^{-3}$) &
($\rm 10^{-4}\,Mpc^{-3}$) &

($\rm 10^{4}\,Mpc^{-3}$) &
($\rm 10^{4}\,Mpc^{-3}$) &
($\rm 10^{4}\,Mpc^{-3}$) 
\\
\hline
\multicolumn{7}{c}{XMM/SDSS}\\ \hline

--18.25 & $1.3 \pm0.7$      &  $0.3\pm 0.3$     &   $1.0\pm0.6$      & $175.8\pm 7.2$ &	 $77.8\pm 4.0$   &   $98.0\pm 6.0$\\  
--18.75 & $1.3\pm 0.5$      &  $1.0\pm 0.5$     &   $0.3\pm0.2$      & $142.9\pm 4.5$ &    $70.0\pm 2.8$   &  $72.9\pm 3.5$  \\
--19.25 & $2.5\pm 0.6$      &  $1.4\pm 0.5$     &   $1.0\pm0.4$      & $97.1\pm 2.7$   &    $50.0\pm 1.7$   &  $47.1\pm 2.0$  \\
--19.75 & $2.3\pm 0.5$      &  $0.8\pm 0.3$     &   $1.2\pm0.4$      & $61.9\pm 1.5$   &    $33.8\pm 1.1$   &  $28.0\pm 1.1$  \\
--20.25 & $2.9 \pm0.7$      &  $2.1\pm 0.6$     &   $0.8\pm0.2$      & $32.8\pm 0.8$   &    $19.3\pm 0.6$   &  $13.5\pm 0.6$  \\
--20.75 & $1.0\pm 0.2$      &  $0.6\pm 0.2$     &   $0.2\pm0.07$    & $12.3\pm0.4$	   &    $7.7\pm 0.3$     &    $4.6\pm 0.5$   \\
--21.25 & $0.6\pm0.2$       &  $0.5\pm 0.2$     &   $0.06\pm0.03$  & $2.7\pm0.2$	   &	  $1.8\pm 0.1$     &    $0.9\pm 0.1$   \\
--21.75 & $0.05\pm0.02$   &  $0.01\pm 0.01$ &   $0.02\pm0.02$  & $0.3\pm0.05$	   &	  $0.20\pm 0.04$ &    $0.1\pm 0.03$  \\ 
--22.25 &  --                       &  --                       &  --                       & $0.03\pm0.02$   &    $0.01\pm 0.01$ &    $0.02\pm 0.01$  \\                      
--22.75 & $0.01\pm0.01$   &  --                       &  --                       & $0.03\pm 0.01$  &    $0.01\pm 0.01$ &    $0.02\pm 0.01$     \\ 
\hline

\multicolumn{7}{c}{NHS}\\ \hline
--18.25 & $1.4 \pm1.4$      &  --                      &   $1.4\pm1.4$      & $128.8\pm 8.2$ &	 $22.6\pm 3.5$   &   $106.1\pm 7.5$\\  
--18.75 & $3.6\pm 3.1$      &  $3.2\pm 3.1$     &   $0.3\pm0.3$      & $106.3\pm 7.3$ &    $29.1\pm 3.9$   &  $77.3\pm 6.3$  \\
--19.25 & $4.7\pm 3.0$      &  $2.2\pm 2.0$     &   $2.4\pm2.2$      & $90.6\pm 6.8$   &    $32.6\pm 4.1$   &  $58.0\pm 5.3$  \\
--19.75 & $1.4\pm 0.6$      &  $1.1\pm 0.5$     &   $0.2\pm0.1$      & $58.2\pm 5.4$   &    $21.6\pm 3.4$   &  $36.6\pm 4.3$  \\
--20.25 & $10.4 \pm5.3$    &  $3.8\pm 2.8$     &   $6.5\pm4.5$      & $31.8\pm 4.2$   &    $16.6\pm 3.1$   &  $15.2\pm 2.8$  \\
--20.75 & $1.1\pm 0.3$      &  $0.4\pm 0.2$     &   $0.4\pm0.2$    & $13.2\pm3.4$	   &    $6.1\pm 1.9$     &    $7.1\pm 2.8$   \\
--21.25 & $0.5\pm0.2$       &  $0.2\pm 0.1$     &   $0.02\pm0.1$  & $3.1\pm1.8$	   &	  $2.1\pm 1.5$     &    $1.0\pm 1.0$   \\
\hline

\multicolumn{7}{c}{AEGIS/GOODS}\\ \hline
--18.25 & $3.2 \pm1.0$      &  $0.4\pm 0.4$     &   $1.2\pm0.9$      & $140.2\pm 7.2$ &	 $9.4\pm 3.2$   &   $130.8\pm 6.5$\\  
--18.75 & $7.8\pm 6.2$      &  $1.0\pm 0.6$     &   $6.5\pm6.1$      & $124.9\pm 4.9$ &    $17.1\pm 2.7$   &  $107.8\pm 4.0$  \\
--19.25 & $2.8\pm 1.5$      &  $0.6\pm0.6$      &   $2.2\pm1.4$      & $105.6\pm 3.5$   &    $25.3\pm 2.3$   &  $80.4\pm 2.6$  \\
--19.75 & $9.6\pm 3.0$      &  $3.6\pm 1.4$     &   $5.0\pm2.6$      & $83.9\pm 2.5$   &    $26.7\pm 1.8$   &  $57.2\pm 1.8$  \\
--20.25 & $20.1 \pm8.7$    &  $14.5\pm 8.5$   &   $4.3\pm1.8$      & $62.5\pm 1.8$   &    $22.3\pm 1.3$   &  $40.2\pm 1.3$  \\
--20.75 & $9.7\pm 2.4$      &  $2.7\pm 1.0$     &   $6.3\pm2.1$      & $42.4\pm1.4$	   &    $18.4\pm 1.0$     &    $24.0\pm 1.0$   \\
--21.25 & $5.1\pm1.1$       &  $1.6\pm 0.4$     &   $2.7\pm1.0$       & $18.3\pm0.9$	   &	  $8.6\pm 0.6$     &    $9.6\pm 0.6$   \\
--21.75 & $1.9\pm0.8$       &  $0.9\pm 0.5$     &   $0.7\pm0.6$       & $5.7\pm0.5$	   &	  $2.7\pm 0.3$ &    $3.0\pm 0.3$  \\ 
--22.25 &  $0.3\pm0.1$      & $0.08\pm0.06$   &   $0.06\pm0.04$   & $0.9\pm0.2$   &    $0.5\pm 0.1$ &    $0.4\pm 0.1$  \\             
--22.75 & $0.4\pm0.3$       &  --                       &   $0.04\pm0.04$   & $0.25\pm 0.09$  &    $0.15\pm 0.07$ &    $0.10\pm 0.06$     \\ 
--23.25 & $0.05\pm0.04$   &  --                       &  --                         & $0.03\pm 0.03$  &    $0.03\pm 0.03$ &    --    \\

\hline

\end{tabular} 
\begin{list}{}{}
\item 
The columns  are: (1): Absolute magnitude $M_{^{0.1}g}$ at the middle of the bin with size 
$\Delta M_{^{0.1}g}=0.5$.; (2):  space density
for all X-ray  AGN; (3) space density for X-ray AGN  in red hosts; (4)
space  density for  X-ray  AGN  in blue  hosts;  (5) space density of galaxies;  (6) 
space density of red galaxies;  (7) space density of blue galaxies.
\end{list}
\end{center}
\end{table*}

\begin{table*}
\caption{Mass Function of AGN and galaxies}\label{tab_MF}
\begin{center} 
\scriptsize
\begin{tabular}{c c c  c c c c}

\hline
$\log M_{star}$ &  
$\phi_{AGN}(ALL)$ &
$\phi_{AGN}(red)$ &
$\phi_{AGN}(blue)$ &

$\phi_{gal}(ALL)$ &
$\phi_{gal}(red)$ &
$\phi_{gal}(blue)$ 
\\

($M_{\odot}$) &
($\rm 10^{-4}\,Mpc^{-3}$) &
($\rm 10^{-4}\,Mpc^{-3}$) &
($\rm 10^{-4}\,Mpc^{-3}$) &

($\rm 10^{4}\,Mpc^{-3}$) &
($\rm 10^{4}\,Mpc^{-3}$) &
($\rm 10^{4}\,Mpc^{-3}$) 
\\
\hline
\multicolumn{7}{c}{XMM/SDSS}\\ \hline

9.1  &   $0.5\pm0.5$ &  --           & $0.5\pm0.5$   &  $412.4\pm 51.3$& $203.5\pm46.7$ & $208.9\pm21.1$ \\
9.3  &  $1.4\pm1.2$  & $0.3\pm0.3$   & $1.1\pm1.1$   &  $242.1\pm 19.4$& $110.2\pm15.9$ & $132.0\pm11.1$ \\
9.5  &  $1.5\pm1.0$  &  --           & $1.5\pm1.0$   &  $243.9\pm 14.3$& $122.8\pm11.6$ & $121.1\pm8.3$  \\
9.7  &  $2.1\pm1.0$  & $0.8\pm0.8$   & $1.3\pm0.6$   &  $263.9\pm 11.2$& $159.4\pm9.6$  & $104.5\pm5.7$  \\
9.9  &  $3.5\pm1.0$  & $1.6\pm0.9$   & $1.9\pm0.6$   &  $224.4\pm 8.0$ & $161.5\pm7.3$  & $62.9\pm 3.4$  \\
10.1 &  $5.4\pm1.4$  & $2.8\pm1.2$   & $2.6\pm0.8$   &  $162.3\pm 5.3$ & $126.3\pm4.9$  & $36.1\pm2.0$   \\
10.3 &  $4.5\pm1.2$  & $2.8\pm1.0$   & $1.8\pm0.6$   &  $113.7\pm 3.5$ & $95.0\pm3.3$   & $18.7\pm1.2$   \\ 
10.5 &  $4.7\pm1.4$  & $4.3\pm1.4$   & $0.4\pm0.1$   &   $66.7\pm 2.1$ & $59.4\pm2.0$   & $7.3\pm0.6$    \\
10.7 &  $1.5\pm0.5$  & $1.2\pm0.5$   & $0.2\pm0.1$   &   $34.6\pm 1.2$ & $31.6\pm1.1$   & $3.0\pm0.4$    \\
10.9 &  $1.8\pm1.1$  & $1.8\pm1.1$   & $0.04\pm0.04$ &   $13.1\pm 0.6$ & $11.9\pm0.6$   & $1.2\pm0.2$    \\
11.1 &  $0.8\pm0.4$  & $0.8\pm0.4$   & --            &    $3.9\pm 0.3$ & $3.5\pm0.3$    & $0.5\pm0.1$    \\
11.3 &    --         & --            & --            &    $0.9\pm 0.2$ & $0.7\pm0.1$    & $0.19\pm0.08$  \\
11.5 &  $0.02\pm0.02$& $0.02\pm0.02$ & --            &    $0.3\pm 0.2$ & $0.3\pm0.1$    & $0.05\pm0.04$  \\
11.7 &    --         & --            & --            &    $0.03\pm0.03$&   --           & $0.03\pm0.03$  \\
\hline

\multicolumn{7}{c}{NHS}\\ \hline
9.1  &   --          &  --           & --            &  $233.7\pm 26.0$& $29.1pm9.4$   & $204.6\pm24.3$ \\
9.3  &  $0.6\pm0.6$  &  --           & $0.6\pm0.6$   &  $153.8\pm 26.0$& $33.5\pm9.7$  & $120.3\pm17.6$ \\
9.5  &  $1.1\pm1.0$  &  --           & $1.1\pm1.0$   &  $178.1\pm 22.9$& $86.6\pm17.1$ & $91.5\pm15.3$  \\
9.7  &  $4.8\pm3.6$  & $0.4\pm0.4$   & $4.4\pm3.6$   &  $142.0\pm 19.9$& $77.9\pm15.4$ & $64.1\pm12.6$  \\
9.9  &  $11.1\pm8.9$ & $2.4\pm2.1$   & $8.8\pm8.7$   &  $100.7\pm 16.0$& $61.0\pm12.5$ & $39.6\pm 9.9$  \\
10.1 &  $7.4\pm5.1$  & $6.9\pm5.1$   & $0.5\pm0.3$   &  $126.2\pm 18.2$& $73.0\pm13.6$ & $53.2\pm12.0$   \\
10.3 &  $8.5\pm7.2$  & $0.8\pm0.4$   & $7.7\pm7.2$   &  $128.5\pm 20.8$& $103.5\pm16.7$& $24.9\pm12.4$   \\ 
10.5 &  $1.8\pm0.9$  & $1.2\pm0.9$   & $0.6\pm0.2$   &   $57.0\pm 12.7$& $51.6\pm12.1$ & $5.4\pm3.8$    \\
10.7 &  $15.1\pm10.1$& $14.9\pm10.1$ & $0.2\pm0.1$   &   $64.6\pm 14.3$& $64.6\pm14.3$ & --    \\
10.9 &  $0.8\pm0.5$  & $0.6\pm0.5$   & $0.2\pm0.2$   &   $42.6\pm 13.1$& $42.6\pm13.1$ & --    \\
11.1 &  --           & --            & --            &    $4.2\pm 4.2$ & $4.2\pm4.2$   & --    \\
11.3 &  $0.07\pm0.07$& $0.07\pm0.07$ & --            &    --           &  --           & --    \\
11.5 &  --           &               & --            &    --           &  --           & --    \\
11.7 &  --           & --            & --            &    --           &  --           & --    \\
\hline

\multicolumn{7}{c}{AEGIS/GOODS}\\ \hline
9.1  &  --           &  --           & --            &  $154.6\pm 11.4$& $2.9\pm1.5$ & $151.7\pm11.3$ \\
9.3  &  $6.7\pm4.4$  &  --           & $6.7\pm4.4$   &  $172.3\pm 18.6$& $8.3\pm2.9$ & $164.0\pm18.4$ \\
9.5  &  $1.8\pm0.9$  & $0.2\pm0.2$   & $1.6\pm0.9$   &  $131.5\pm  8.7$& $9.7\pm2.8$ & $121.8\pm8.3$  \\
9.7  &  $10.2\pm4.7$ & $1.1\pm0.8$   & $9.1\pm4.6$   &  $128.1\pm  9.5$& $26.6\pm6.8$  & $101.4\pm6.6$  \\
9.9  &  $9.6\pm3.5$  & $2.8\pm1.4$   & $6.9\pm3.2$   &  $106.7\pm 8.3$ & $32.2\pm6.2$  & $74.5\pm 5.5$  \\
10.1 &  $25.8\pm16.4$& $3.5\pm2.5$   & $22.3\pm16.2$ &  $ 85.4\pm 6.3$ & $28.5\pm4.5$  & $57.0\pm4.5$   \\
10.3 &  $18.0\pm5.9$ & $5.3\pm1.9$   & $12.8\pm5.6$  &  $ 72.1\pm 5.6$ & $39.0\pm4.4$   & $33.1\pm3.4$   \\ 
10.5 &  $19.8\pm4.6$ & $10.6\pm3.7$  & $9.1\pm2.7$   &   $72.7\pm 5.4$ & $42.0\pm4.5$   & $30.7\pm3.1$    \\
10.7 &  $31.0\pm21.2$& $30.1\pm21.2$ & $0.9\pm0.3$   &   $47.1\pm 3.7$ & $31.3\pm3.2$   & $15.9\pm1.8$    \\
10.9 &  $1.7\pm0.6$  & $1.1\pm1.1$   & $0.6\pm0.5$   &   $25.8\pm 2.7$ & $19.3\pm2.3$   & $6.4\pm1.3$    \\
11.1 &  $2.6\pm1.1$  & $2.4\pm1.1$   & $0.2\pm0.2$   &    $17.1\pm 2.3$& $12.9\pm1.9$   & $4.2\pm1.4$    \\
11.3 &  $0.6\pm0.3$  & $0.4\pm0.3$   & $0.2\pm0.2$   &    $7.8\pm 1.4$ & $6.7\pm1.3$    & $1.1\pm0.6$  \\
11.5 &  $1.9\pm1.3$  & $1.9\pm1.3$   & --            &    $1.5\pm 0.6$ & $1.3\pm0.6$    & $0.2\pm0.2$  \\
11.7 &    --         & --            & --            &    $0.8\pm0.4$  & $0.6\pm0.4$    & $0.2\pm0.2$  \\
\hline

\hline

\end{tabular} 
\begin{list}{}{}
\item 
The columns  are: (1): Stellar mass at the middle of the bin with size 
$\Delta\log M_{star}=0.2$; (2):  space density 
for all X-ray  AGN; (3) space density for X-ray AGN  in red hosts; (4)
space  density for  X-ray  AGN  in blue  hosts;  (5) space density of galaxies;  (6) 
space density of red galaxies;  (7) space density of blue galaxies.
\end{list}
\end{center}
\end{table*}

\section{Acknowledgments}
AG acknowledges  financial support from  the Marie-Curie Reintegration
Grant  PERG03-GA-2008-230644.  Based  on  observations made  with  ESO
Telescopes at  the La Silla and Paranal  Observatories under programme
IDs 078.B-0623A and 080.B-0409A. Funding for the DEEP2 Galaxy Redshift
Survey  has   been  provided  in  part  by   NSF  grants  AST95-09298,
AST-0071048, AST-0071198, AST-0507428, and AST-0507483 as well as NASA
LTSA grant NNG04GC89G. Funding for the Sloan Digital Sky Survey (SDSS)
has   been  provided  by   the  Alfred   P.   Sloan   Foundation,  the
Participating  Institutions,   the  National  Aeronautics   and  Space
Administration, the National  Science Foundation, the U.S.  Department
of Energy,  the Japanese Monbukagakusho,  and the Max  Planck Society.
The SDSS Web site is http://www.sdss.org/.  The SDSS is managed by the
Astrophysical   Research  Consortium   (ARC)  for   the  Participating
Institutions.   The Participating Institutions  are The  University of
Chicago,  Fermilab,  the  Institute  for  Advanced  Study,  the  Japan
Participation Group, The Johns Hopkins University, Los Alamos National
Laboratory,  the   Max-Planck-Institute  for  Astronomy   (MPIA),  the
Max-Planck-Institute   for  Astrophysics   (MPA),  New   Mexico  State
University, University of Pittsburgh, Princeton University, the United
States Naval Observatory, and the University of Washington.

\bibliography{mybib}{}

\begin{thebibliography}{92}
\expandafter\ifx\csname natexlab\endcsname\relax\def\natexlab#1{#1}\fi

\bibitem[{{Abazajian} {et~al.}(2009)}]{Abazajian2009}
{Abazajian} K.~N., {et~al.}, 2009, ApJS, 182, 543

\bibitem[{{Aird} {et~al.}(2010)}]{Aird2010}
{Aird} J., {et~al.}, 2010, MNRAS, 401, 2531

\bibitem[{{Allevato} {et~al.}(2011)}]{Allevato2011}
{Allevato} V., {et~al.}, 2011, ArXiv 1105.0520

\bibitem[{{Balestra} {et~al.}(2010){Balestra}, {Mainieri}, {Popesso},
  {Dickinson}, {Nonino}, {Rosati}, {Teimoorinia}, {Vanzella}, {Cristiani},
  {Cesarsky}, {Fosbury}, {Kuntschner}, \& {Rettura}}]{Balestra2010}
{Balestra} I., {Mainieri} V., {Popesso} P., {Dickinson} M., {Nonino} M.,
  {Rosati} P., {Teimoorinia} H., {Vanzella} E., {Cristiani} S., {Cesarsky} C.,
  {Fosbury} R.~A.~E., {Kuntschner} H., {Rettura} A., 2010, A\&A, 512, A12+

\bibitem[{{Barger} {et~al.}(2003){Barger}, {Cowie}, {Capak}, {Alexander},
  {Bauer}, {Fernandez}, {Brandt}, {Garmire}, \& {Hornschemeier}}]{Barger2003}
{Barger} A.~J., {Cowie} L.~L., {Capak} P., {Alexander} D.~M., {Bauer} F.~E.,
  {Fernandez} E., {Brandt} W.~N., {Garmire} G.~P., {Hornschemeier} A.~E., 2003,
  AJ, 126, 632

\bibitem[{{Barger} {et~al.}(2005){Barger}, {Cowie}, {Mushotzky}, {Yang},
  {Wang}, {Steffen}, \& {Capak}}]{Barger2005}
{Barger} A.~J., {Cowie} L.~L., {Mushotzky} R.~F., {Yang} Y., {Wang} W.-H.,
  {Steffen} A.~T., {Capak} P., 2005, AJ, 129, 578

\bibitem[{{Benson} {et~al.}(2003){Benson}, {Bower}, {Frenk}, {Lacey}, {Baugh},
  \& {Cole}}]{Benson2003}
{Benson} A.~J., {Bower} R.~G., {Frenk} C.~S., {Lacey} C.~G., {Baugh} C.~M.,
  {Cole} S., 2003, ApJ, 599, 38

\bibitem[{{Blanton}(2006)}]{Blanton2006}
{Blanton} M.~R., 2006, ApJ, 648, 268

\bibitem[{{Blanton} \& {Roweis}(2007)}]{Blanton_Roweis2007}
{Blanton} M.~R., {Roweis} S., 2007, AJ, 133, 734

\bibitem[{{Blanton} {et~al.}(2005){Blanton}, {Schlegel}, {Strauss},
  {Brinkmann}, {Finkbeiner}, {Fukugita}, {Gunn}, {Hogg}, {Ivezi{\'c}}, {Knapp},
  {Lupton}, {Munn}, {Schneider}, {Tegmark}, \& {Zehavi}}]{Blanton2005}
{Blanton} M.~R., {Schlegel} D.~J., {Strauss} M.~A., {Brinkmann} J.,
  {Finkbeiner} D., {Fukugita} M., {Gunn} J.~E., {Hogg} D.~W., {Ivezi{\'c}} {\v
  Z}., {Knapp} G.~R., {Lupton} R.~H., {Munn} J.~A., {Schneider} D.~P.,
  {Tegmark} M., {Zehavi} I., 2005, AJ, 129, 2562

\bibitem[{{Borch} {et~al.}(2006){Borch}, {Meisenheimer}, {Bell}, {Rix}, {Wolf},
  {Dye}, {Kleinheinrich}, {Kovacs}, \& {Wisotzki}}]{Borch2006}
{Borch} A., {Meisenheimer} K., {Bell} E.~F., {Rix} H., {Wolf} C., {Dye} S.,
  {Kleinheinrich} M., {Kovacs} Z., {Wisotzki} L., 2006, A\&A, 453, 869

\bibitem[{{Bundy} {et~al.}(2008)}]{Bundy2008}
{Bundy} K., {et~al.}, 2008, ApJ, 681, 931

\bibitem[{{Cardamone} {et~al.}(2010){Cardamone}, {Urry}, {Schawinski},
  {Treister}, {Brammer}, \& {Gawiser}}]{Cardamone2010}
{Cardamone} C.~N., {Urry} C.~M., {Schawinski} K., {Treister} E., {Brammer} G.,
  {Gawiser} E., 2010, ApJ, 721, L38

\bibitem[{{Cattaneo} {et~al.}(2007){Cattaneo}, {Blaizot}, {Weinberg}, {Kere{\v
  s}}, {Colombi}, {Dav{\'e}}, {Devriendt}, {Guiderdoni}, \&
  {Katz}}]{Cattaneo2007}
{Cattaneo} A., {Blaizot} J., {Weinberg} D.~H., {Kere{\v s}} D., {Colombi} S.,
  {Dav{\'e}} R., {Devriendt} J., {Guiderdoni} B., {Katz} N., 2007, MNRAS, 377,
  63

\bibitem[{{Cattaneo} {et~al.}(2009)}]{Cattaneo2009}
{Cattaneo} A., {et~al.}, 2009, Nature, 460, 213

\bibitem[{{Cen}(2011)}]{Cen2011}
{Cen} R., 2011, arXiv:1102.0262

\bibitem[{{Ciotti} \& {Ostriker}(1997)}]{Ciotti_Ostriker1997}
{Ciotti} L., {Ostriker} J.~P., 1997, ApJ, 487, L105+

\bibitem[{{Ciotti} \& {Ostriker}(2007)}]{Ciotti_Ostriker2007}
---, 2007, ApJ, 665, 1038

\bibitem[{{Cisternas} {et~al.}(2011)}]{Cisternas2011}
{Cisternas} M., {et~al.}, 2011, ApJ, 726, 57

\bibitem[{{Coil} {et~al.}(2009){Coil}, {Georgakakis}, {Newman}, {Cooper},
  {Croton}, {Davis}, {Koo}, {Laird}, {Nandra}, {Weiner}, {Willmer}, \&
  {Yan}}]{Coil2009}
{Coil} A.~L., {Georgakakis} A., {Newman} J.~A., {Cooper} M.~C., {Croton} D.,
  {Davis} M., {Koo} D.~C., {Laird} E.~S., {Nandra} K., {Weiner} B.~J.,
  {Willmer} C.~N.~A., {Yan} R., 2009, ApJ, 701, 1484

\bibitem[{{Coil} {et~al.}(2004){Coil}, {Newman}, {Kaiser}, {Davis}, {Ma},
  {Kocevski}, \& {Koo}}]{Coil2004}
{Coil} A.~L., {Newman} J.~A., {Kaiser} N., {Davis} M., {Ma} C., {Kocevski}
  D.~D., {Koo} D.~C., 2004, ApJ, 617, 765

\bibitem[{{Cowie} {et~al.}(2003){Cowie}, {Barger}, {Bautz}, {Brandt}, \&
  {Garmire}}]{Cowie2003}
{Cowie} L.~L., {Barger} A.~J., {Bautz} M.~W., {Brandt} W.~N., {Garmire} G.~P.,
  2003, ApJ, 584, L57

\bibitem[{{Croton} {et~al.}(2006)}]{Croton2006}
{Croton} D.~J., {et~al.}, 2006, MNRAS, 365, 11

\bibitem[{{Davis} {et~al.}(2003)}]{Davis2003}
{Davis} M., {et~al.}, 2003, in Society of Photo-Optical Instrumentation
  Engineers (SPIE) Conference Series, {Guhathakurta} P., ed., Vol. 4834, pp.
  161--172

\bibitem[{{Davis} {et~al.}(2007)}]{Davis2007}
---, 2007, ApJ, 660, L1

\bibitem[{{Di Matteo} {et~al.}(2005){Di Matteo}, {Springel}, \&
  {Hernquist}}]{DiMatteo2005}
{Di Matteo} T., {Springel} V., {Hernquist} L., 2005, Nature, 433, 604

\bibitem[{{Dickinson} {et~al.}(2004)}]{Dickinson2004}
{Dickinson} M., {et~al.}, 2004, ApJ, 600, L99

\bibitem[{{Faber} {et~al.}(2003)}]{Faber2003}
{Faber} S.~M., {et~al.}, 2003, in Society of Photo-Optical Instrumentation
  Engineers (SPIE) Conference, Vol. 4841, Instrument Design and Performance for
  Optical/Infrared Ground-based Telescopes, {Iye} M., {Moorwood} A.~F.~M.,
  eds., pp. 1657--1669

\bibitem[{{Fanidakis} {et~al.}(2010){Fanidakis}, {Baugh}, {Benson}, {Bower},
  {Cole}, {Done}, {Frenk}, {Hickox}, {Lacey}, \& {Lagos}}]{Fanidakis2011}
{Fanidakis} N., {Baugh} C.~M., {Benson} A.~J., {Bower} R.~G., {Cole} S., {Done}
  C., {Frenk} C.~S., {Hickox} R.~C., {Lacey} C., {Lagos} C.~d.~P., 2010,
  arXiv-1011.5222

\bibitem[{{Ferrarese} \& {Merritt}(2000)}]{Ferrarese2000}
{Ferrarese} L., {Merritt} D., 2000, ApJ, 539, L9

\bibitem[{{Fontanot} {et~al.}(2007{\natexlab{a}}){Fontanot}, {Cristiani},
  {Monaco}, {Nonino}, {Vanzella}, {Brandt}, {Grazian}, \& {Mao}}]{Fontanot2007}
{Fontanot} F., {Cristiani} S., {Monaco} P., {Nonino} M., {Vanzella} E.,
  {Brandt} W.~N., {Grazian} A., {Mao} J., 2007{\natexlab{a}}, A\&A, 461, 39

\bibitem[{{Fontanot} {et~al.}(2007{\natexlab{b}}){Fontanot}, {Monaco}, {Silva},
  \& {Grazian}}]{Fontanot2007_lf}
{Fontanot} F., {Monaco} P., {Silva} L., {Grazian} A., 2007{\natexlab{b}},
  MNRAS, 382, 903

\bibitem[{{Gabor} {et~al.}(2009)}]{Gabor2009}
{Gabor} J.~M., {et~al.}, 2009, ApJ, 691, 705

\bibitem[{{Gebhardt} {et~al.}(2000){Gebhardt}, {Bender}, {Bower}, {Dressler},
  {Faber}, {Filippenko}, {Green}, {Grillmair}, {Ho}, {Kormendy}, {Lauer},
  {Magorrian}, {Pinkney}, {Richstone}, \& {Tremaine}}]{Gebhardt2000}
{Gebhardt} K., {Bender} R., {Bower} G., {Dressler} A., {Faber} S.~M.,
  {Filippenko} A.~V., {Green} R., {Grillmair} C., {Ho} L.~C., {Kormendy} J.,
  {Lauer} T.~R., {Magorrian} J., {Pinkney} J., {Richstone} D., {Tremaine} S.,
  2000, ApJ, 539, L13

\bibitem[{{Georgakakis} {et~al.}(2008){Georgakakis}, {Gerke}, {Nandra},
  {Laird}, {Coil}, {Cooper}, \& {Newman}}]{Georgakakis2008_groups}
{Georgakakis} A., {Gerke} B.~F., {Nandra} K., {Laird} E.~S., {Coil} A.~L.,
  {Cooper} M.~C., {Newman} J.~A., 2008, MNRAS, 391, 183

\bibitem[{{Georgakakis} \& {Nandra}(2011)}]{Georgakakis_Nandra2011}
{Georgakakis} A., {Nandra} K., 2011, ArXiv: 1101.4943

\bibitem[{{Georgakakis} {et~al.}(2006)}]{Georgakakis2006}
{Georgakakis} A., {et~al.}, 2006, MNRAS, 371, 221

\bibitem[{{Georgakakis} {et~al.}(2009)}]{Georgakakis2009}
---, 2009, MNRAS, 397, 623

\bibitem[{{Georgantopoulos} \& {Akylas}(2010)}]{Georgantopoulos_Akylas2010}
{Georgantopoulos} I., {Akylas} A., 2010, A\&A, 509, A38+

\bibitem[{{Georgantopoulos} {et~al.}(2005){Georgantopoulos}, {Georgakakis}, \&
  {Koulouridis}}]{Georgantopoulos2005}
{Georgantopoulos} I., {Georgakakis} A., {Koulouridis} E., 2005, MNRAS, 360, 782

\bibitem[{{Green} {et~al.}(2009){Green}, {Aldcroft}, {Richards}, {Barkhouse},
  {Constantin}, {Haggard}, {Karovska}, {Kim}, {Kim}, {Vikhlinin}, {Anderson},
  {Mossman}, {Kashyap}, {Myers}, {Silverman}, {Wilkes}, \&
  {Tananbaum}}]{Green2009}
{Green} P.~J., {Aldcroft} T.~L., {Richards} G.~T., {Barkhouse} W.~A.,
  {Constantin} A., {Haggard} D., {Karovska} M., {Kim} D., {Kim} M., {Vikhlinin}
  A., {Anderson} S.~F., {Mossman} A., {Kashyap} V., {Myers} A.~C., {Silverman}
  J.~D., {Wilkes} B.~J., {Tananbaum} H., 2009, ApJ, 690, 644

\bibitem[{{Haggard} {et~al.}(2010){Haggard}, {Green}, {Anderson}, {Constantin},
  {Aldcroft}, {Kim}, \& {Barkhouse}}]{Haggard2010}
{Haggard} D., {Green} P.~J., {Anderson} S.~F., {Constantin} A., {Aldcroft}
  T.~L., {Kim} D., {Barkhouse} W.~A., 2010, ApJ, 723, 1447

\bibitem[{{Hasinger}(2008)}]{Hasinger2008}
{Hasinger} G., 2008, A\&A, 490, 905

\bibitem[{{Hickox} {et~al.}(2009)}]{Hickox2009}
{Hickox} R.~C., {et~al.}, 2009, ApJ, 696, 891

\bibitem[{{Hopkins}(2004)}]{Hopkins_A2004}
{Hopkins} A.~M., 2004, ApJ, 615, 209

\bibitem[{{Hopkins} \& {Hernquist}(2006)}]{Hopkins_Hernquist2006}
{Hopkins} P.~F., {Hernquist} L., 2006, ApJS, 166, 1

\bibitem[{{Hopkins} {et~al.}(2008){Hopkins}, {Hernquist}, {Cox}, \& {Kere{\v
  s}}}]{Hopkins2008_sam}
{Hopkins} P.~F., {Hernquist} L., {Cox} T.~J., {Kere{\v s}} D., 2008, ApJS, 175,
  356

\bibitem[{{Hornschemeier} {et~al.}(2003){Hornschemeier}, {Bauer}, {Alexander},
  {Brandt}, {Sargent}, {Bautz}, {Conselice}, {Garmire}, {Schneider}, \&
  {Wilson}}]{Hornschemeier2003}
{Hornschemeier} A.~E., {Bauer} F.~E., {Alexander} D.~M., {Brandt} W.~N.,
  {Sargent} W.~L.~W., {Bautz} M.~W., {Conselice} C., {Garmire} G.~P.,
  {Schneider} D.~P., {Wilson} G., 2003, AJ, 126, 575

\bibitem[{{Juneau} {et~al.}(2011){Juneau}, {Dickinson}, {Alexander}, \&
  {Salim}}]{Juneau2011}
{Juneau} S., {Dickinson} M., {Alexander} D.~M., {Salim} S., 2011, ArXiv
  1105.3194

\bibitem[{{Kauffmann} \& {Heckman}(2009)}]{Kauffmann_Heckman2009}
{Kauffmann} G., {Heckman} T.~M., 2009, MNRAS, 397, 135

\bibitem[{{Kauffmann} {et~al.}(2003){Kauffmann}, {Heckman}, {Tremonti},
  {Brinchmann}, {Charlot}, {White}, {Ridgway}, {Brinkmann}, {Fukugita}, {Hall},
  {Ivezi{\'c}}, {Richards}, \& {Schneider}}]{Kauffmann2003}
{Kauffmann} G., {Heckman} T.~M., {Tremonti} C., {Brinchmann} J., {Charlot} S.,
  {White} S.~D.~M., {Ridgway} S.~E., {Brinkmann} J., {Fukugita} M., {Hall}
  P.~B., {Ivezi{\'c}} {\v Z}., {Richards} G.~T., {Schneider} D.~P., 2003,
  MNRAS, 346, 1055

\bibitem[{{Kauffmann} {et~al.}(2004){Kauffmann}, {White}, {Heckman},
  {M{\'e}nard}, {Brinchmann}, {Charlot}, {Tremonti}, \&
  {Brinkmann}}]{Kauffmann2004}
{Kauffmann} G., {White} S.~D.~M., {Heckman} T.~M., {M{\'e}nard} B.,
  {Brinchmann} J., {Charlot} S., {Tremonti} C., {Brinkmann} J., 2004, MNRAS,
  353, 713

\bibitem[{{Kenter} {et~al.}(2005)}]{Kenter2005}
{Kenter} A., {et~al.}, 2005, ApJS, 161, 9

\bibitem[{{Laird} {et~al.}(2009)}]{Laird2009}
{Laird} E.~S., {et~al.}, 2009, ApJS, 180, 102

\bibitem[{{Le F{\`e}vre} {et~al.}(2004)}]{LeFevre2004}
{Le F{\`e}vre} O., {et~al.}, 2004, A\&A, 428, 1043

\bibitem[{{Lin} {et~al.}(1999){Lin}, {Yee}, {Carlberg}, {Morris}, {Sawicki},
  {Patton}, {Wirth}, \& {Shepherd}}]{Lin1999}
{Lin} H., {Yee} H.~K.~C., {Carlberg} R.~G., {Morris} S.~L., {Sawicki} M.,
  {Patton} D.~R., {Wirth} G., {Shepherd} C.~W., 1999, ApJ, 518, 533

\bibitem[{{Lin} {et~al.}(2010)}]{Lin2010}
{Lin} L., {et~al.}, 2010, ApJ, 718, 1158

\bibitem[{{Lutz} {et~al.}(2010)}]{Lutz2010}
{Lutz} D., {et~al.}, 2010, ApJ, 712, 1287

\bibitem[{{Mignoli} {et~al.}(2005)}]{Mignoli2005}
{Mignoli} M., {et~al.}, 2005, A\&A, 437, 883

\bibitem[{{Monaco} {et~al.}(2007){Monaco}, {Fontanot}, \&
  {Taffoni}}]{Monaco2007}
{Monaco} P., {Fontanot} F., {Taffoni} G., 2007, MNRAS, 375, 1189

\bibitem[{{Morrison} \& {McCammon}(1983)}]{Morrison1983}
{Morrison} R., {McCammon} D., 1983, ApJ, 270, 119

\bibitem[{{Nandra} \& {Pounds}(1994)}]{Nandra1994}
{Nandra} K., {Pounds} K.~A., 1994, MNRAS, 268, 405

\bibitem[{{Netzer} {et~al.}(2006){Netzer}, {Mainieri}, {Rosati}, \&
  {Trakhtenbrot}}]{Netzer2006}
{Netzer} H., {Mainieri} V., {Rosati} P., {Trakhtenbrot} B., 2006, A\&A, 453,
  525

\bibitem[{{Padmanabhan} {et~al.}(2008)}]{Padmanabhan2008}
{Padmanabhan} N., {et~al.}, 2008, ApJ, 674, 1217

\bibitem[{{P{\'e}rez-Gonz{\'a}lez} {et~al.}(2008)}]{Perez2008}
{P{\'e}rez-Gonz{\'a}lez} P.~G., {et~al.}, 2008, ApJ, 675, 234

\bibitem[{{Pierce} {et~al.}(2007){Pierce}, {Lotz}, {Laird}, {Lin}, {Nandra},
  {Primack}, {Faber}, {Barmby}, {Park}, {Willner}, {Gwyn}, {Koo}, {Coil},
  {Cooper}, {Georgakakis}, {Koekemoer}, {Noeske}, {Weiner}, \&
  {Willmer}}]{Pierce2007}
{Pierce} C.~M., {Lotz} J.~M., {Laird} E.~S., {Lin} L., {Nandra} K., {Primack}
  J.~R., {Faber} S.~M., {Barmby} P., {Park} S.~Q., {Willner} S.~P., {Gwyn} S.,
  {Koo} D.~C., {Coil} A.~L., {Cooper} M.~C., {Georgakakis} A., {Koekemoer}
  A.~M., {Noeske} K.~G., {Weiner} B.~J., {Willmer} C.~N.~A., 2007, ApJ, 660,
  L19

\bibitem[{{Pierce} {et~al.}(2010)}]{Pierce2010}
{Pierce} C.~M., {et~al.}, 2010, MNRAS, 408, 139

\bibitem[{{Popesso} {et~al.}(2008)}]{Popesso2008}
{Popesso} P., {et~al.}, 2008, ArXiv0802.2930

\bibitem[{{Popesso} {et~al.}(2009)}]{Popesso2009}
---, 2009, A\&A, 494, 443

\bibitem[{{Sarzi} {et~al.}(2010)}]{Sarzi2010}
{Sarzi} M., {et~al.}, 2010, MNRAS, 402, 2187

\bibitem[{{Schlegel} {et~al.}(1998){Schlegel}, {Finkbeiner}, \&
  {Davis}}]{Schlegel1998}
{Schlegel} D.~J., {Finkbeiner} D.~P., {Davis} M., 1998, ApJ, 500, 525

\bibitem[{{Schmidt}(1968)}]{Schmidt1968}
{Schmidt} M., 1968, ApJ, 151, 393

\bibitem[{{Scoville} {et~al.}(2007){Scoville}, {Aussel}, {Brusa}, {Capak},
  {Carollo}, {Elvis}, {Giavalisco}, {Guzzo}, {Hasinger}, {Impey}, {Kneib},
  {LeFevre}, {Lilly}, {Mobasher}, {Renzini}, {Rich}, {Sanders}, {Schinnerer},
  {Schminovich}, {Shopbell}, {Taniguchi}, \& {Tyson}}]{Scoville2007}
{Scoville} N., {Aussel} H., {Brusa} M., {Capak} P., {Carollo} C.~M., {Elvis}
  M., {Giavalisco} M., {Guzzo} L., {Hasinger} G., {Impey} C., {Kneib} J.-P.,
  {LeFevre} O., {Lilly} S.~J., {Mobasher} B., {Renzini} A., {Rich} R.~M.,
  {Sanders} D.~B., {Schinnerer} E., {Schminovich} D., {Shopbell} P.,
  {Taniguchi} Y., {Tyson} N.~D., 2007, ApJS, 172, 1

\bibitem[{{Shao} {et~al.}(2010)}]{Shao2010}
{Shao} L., {et~al.}, 2010, A\&A, 518, L26+

\bibitem[{{Shi} {et~al.}(2008){Shi}, {Rieke}, {Donley}, {Cooper}, {Willmer}, \&
  {Kirby}}]{Shi2008}
{Shi} Y., {Rieke} G., {Donley} J., {Cooper} M., {Willmer} C., {Kirby} E., 2008,
  ApJ, 688, 794

\bibitem[{{Silverman} {et~al.}(2009{\natexlab{a}})}]{Silverman2009_SFR}
{Silverman} J.~D., {et~al.}, 2009{\natexlab{a}}, ApJ, 696, 396

\bibitem[{{Silverman} {et~al.}(2009{\natexlab{b}})}]{Silverman2009_env}
---, 2009{\natexlab{b}}, ApJ, 695, 171

\bibitem[{{Somerville} {et~al.}(2008){Somerville}, {Hopkins}, {Cox},
  {Robertson}, \& {Hernquist}}]{Somerville2008}
{Somerville} R.~S., {Hopkins} P.~F., {Cox} T.~J., {Robertson} B.~E.,
  {Hernquist} L., 2008, ArXiv0808.1227

\bibitem[{{Stanway} {et~al.}(2004{\natexlab{a}}){Stanway}, {Bunker}, {McMahon},
  {Ellis}, {Treu}, \& {McCarthy}}]{Stanway2004a}
{Stanway} E.~R., {Bunker} A.~J., {McMahon} R.~G., {Ellis} R.~S., {Treu} T.,
  {McCarthy} P.~J., 2004{\natexlab{a}}, ApJ, 607, 704

\bibitem[{{Stanway} {et~al.}(2004{\natexlab{b}})}]{Stanway2004b}
{Stanway} E.~R., {et~al.}, 2004{\natexlab{b}}, ApJ, 604, L13

\bibitem[{{Strauss} {et~al.}(2002)}]{Strauss2002}
{Strauss} M.~A., {et~al.}, 2002, AJ, 124, 1810

\bibitem[{{Szokoly} {et~al.}(2004)}]{Szokoly2004}
{Szokoly} G.~P., {et~al.}, 2004, ApJS, 155, 271

\bibitem[{{Vanzella} {et~al.}(2005)}]{Vanzella2005}
{Vanzella} E., {et~al.}, 2005, A\&A, 434, 53

\bibitem[{{Vanzella} {et~al.}(2008)}]{Vanzella2008}
---, 2008, A\&A, 478, 83

\bibitem[{{Wang} {et~al.}(2007){Wang}, {Li}, {Kauffman}, \& {De
  Lucia}}]{Wang2007}
{Wang} L., {Li} C., {Kauffman} G., {De Lucia} G., 2007, MNRAS, 377, 1419

\bibitem[{{Weiner} {et~al.}(2005)}]{Weiner2005}
{Weiner} B.~J., {et~al.}, 2005, ApJ, 620, 595

\bibitem[{{Willmer} {et~al.}(2006)}]{Willmer2006}
{Willmer} C.~N.~A., {et~al.}, 2006, ApJ, 647, 853

\bibitem[{{Wirth} {et~al.}(2004)}]{Wirth2004}
{Wirth} G.~D., {et~al.}, 2004, AJ, 127, 3121

\bibitem[{{Xue} {et~al.}(2010){Xue}, {Brandt}, {Luo}, {Rafferty}, {Alexander},
  {Bauer}, {Lehmer}, {Schneider}, \& {Silverman}}]{Xue2010}
{Xue} Y.~Q., {Brandt} W.~N., {Luo} B., {Rafferty} D.~A., {Alexander} D.~M.,
  {Bauer} F.~E., {Lehmer} B.~D., {Schneider} D.~P., {Silverman} J.~D., 2010,
  ApJ, 720, 368

\bibitem[{{Yan} {et~al.}(2011)}]{Yan2011}
{Yan} R., {et~al.}, 2011, ApJ, 728, 38

\bibitem[{{Zheng} {et~al.}(2009){Zheng}, {Bell}, {Somerville}, {Rix}, {Jahnke},
  {Fontanot}, {Rieke}, {Schiminovich}, \& {Meisenheimer}}]{Zheng2009}
{Zheng} X.~Z., {Bell} E.~F., {Somerville} R.~S., {Rix} H., {Jahnke} K.,
  {Fontanot} F., {Rieke} G.~H., {Schiminovich} D., {Meisenheimer} K., 2009,
  ApJ, 707, 1566

\bibitem[{{Zibetti} {et~al.}(2009){Zibetti}, {Charlot}, \& {Rix}}]{zibetti2009}
{Zibetti} S., {Charlot} S., {Rix} H., 2009, MNRAS, 400, 1181

\end{thebibliography}
\bibliographystyle{mn2e}

\end{document}